%% file: GNBC.tex
\documentclass[]{jfm} 

\usepackage{graphicx}
\usepackage{newtxtext}
\usepackage{newtxmath}
\usepackage{natbib}
\usepackage{hyperref}
\hypersetup{
    colorlinks = true,
    urlcolor   = blue,
    citecolor  = black,
}

\newcommand{\RomanNumeralCaps}[1]
\linenumbers

\usepackage{subfig}
\usepackage{placeins}

\title{A consistent treatment of dynamic contact angles in the sharp-interface framework with the generalized Navier boundary condition}

\author{Tomas Fullana\aff{1,2},
  Yash Kulkarni\aff{1},
  Mathis Fricke\aff{3},
  Stéphane Popinet\aff{1},
  Shahriar Afkhami\aff{4},
  Dieter Bothe\aff{3},
  \and Stéphane Zaleski\aff{1,5}}

\affiliation{\aff{1}Sorbonne Université and CNRS, Institut Jean Le Rond d’Alembert UMR 7190, F-75005 Paris, France
\aff{2}Laboratory of Fluid Mechanics and Instabilities, EPFL, Lausanne CH-1015, Switzerland
\aff{3}Department of Mathematics, TU Darmstadt, Schlossgartenstra\ss e 7, 64289 Darmstadt, Germany
\aff{4}Department of Mathematical Sciences, New Jersey Institute of Technology, Newark, NJ, USA 07102
\aff{5}Institut Universitaire de France, Paris, France}

\DeclareMathSymbol{\shortminus}{\mathbin}{AMSa}{"39}
\usepackage{algorithmic}
\usepackage[lined,boxed,ruled,commentsnumbered]{algorithm2e}
\usepackage{dsfont}
\usepackage{gensymb}
\usepackage{booktabs} 

\newcommand{\inproduct}[2]{\left\langle #1 , #2 \right\rangle}
\newcommand{\jump}[1]{\left\llbracket #1 \right\rrbracket}
\newcommand{\nsigma}{\mathbf{n}_\Sigma}
\newcommand{\ddt}[1]{\frac{d #1}{dt}}

\newcommand{\transpose}{\mathsf{T}} 
\newcommand{\ndomega}{\mathbf{n}_{\partial\Omega}}
\newcommand{\ngamma}{\mathbf{n}_\Gamma}
\newcommand{\tgamma}{\mathbf{t}_\Gamma} 
\newcommand{\clspeed}{U_{\text{cl}}}
\newcommand{\normalspeed}{V_\Sigma}
\newcommand{\sigmawet}{\sigma_{\text{w}}}
\newcommand{\visc}{\eta}
\newcommand{\mat}[1]{\mathbf{#1}} 

\DeclareMathOperator{\cawall}{Ca} 
\DeclareMathOperator{\caloc}{Ca_{loc}} 
\DeclareMathOperator{\cacl}{Ca_{cl}} 
\DeclareMathOperator{\catr}{Ca_{tr}} 

\newcommand{\uwall}{U_w} 
\newcommand{\vuwall}{\mathbf{U}_w} 

\DeclareMathOperator{\thetaeq}{\theta_e} 
\DeclareMathOperator{\thetad}{\theta_d} 
\DeclareMathOperator{\thetasteady}{\theta_{s}} 
\DeclareMathOperator{\thetaabove}{\theta_{a}} 
\newcommand{\dthetad}{{\dot{\theta}}_{\text{d}}} 

\DeclareMathOperator{\thetaan}{\theta_{an}} 
\DeclareMathOperator{\kappaan}{\kappa_{an}}

\DeclareMathOperator{\caplength}{\mathit{l}_{c}} 

\DeclareMathOperator{\dirac}{\mathit{f}  \left( \dfrac{\mathit{x} }{\varepsilon} \right)}

\newcommand{\RR}{\mathds{R}} 

\usepackage{xcolor}

\usepackage{nicefrac}

\definecolor{revone}{RGB}{220, 50, 47}     
\definecolor{revtwo}{RGB}{38, 139, 210}    
\definecolor{revthree}{RGB}{133, 153, 0}   
\definecolor{revgen}{RGB}{128, 0, 128}     

\newcommand{\revone}[1]{\textcolor{black}{#1}}     
\newcommand{\revthree}[1]{\textcolor{black}{#1}} 
\newcommand{\revgen}[1]{\textcolor{black}{#1}}     
\newcommand{\secondrev}[1]{\textcolor{black}{#1}}     

\begin{document}
\maketitle

\begin{abstract}
In this work, we revisit the Generalized Navier Boundary condition (GNBC) introduced by Qian et al.\ in the sharp interface Volume-of-Fluid context. We replace the singular uncompensated Young stress by a smooth function with a characteristic width \revgen{$\varepsilon > 0$} that is understood as a physical parameter of the model. Therefore, we call the model the ``Contact Region GNBC'' (CR-GNBC). We show that the model is consistent with the fundamental kinematics of the contact angle transport described by Fricke, Köhne and Bothe. We implement the model in the geometrical Volume-of-Fluid solver Basilisk using a ``free angle'' approach. This means that the dynamic contact angle is not prescribed but reconstructed from the interface geometry and subsequently applied as an input parameter to compute the uncompensated Young stress. We couple this approach to the two-phase Navier Stokes solver and study the withdrawing tape problem with a receding contact line. It is shown that the model \revgen{allows for grid-independent solutions} and leads to a full regularization of the singularity at the moving contact line, which is in accordance with the thin-film equation subject to this boundary condition. In particular, it is shown that the curvature at the moving contact line is finite and mesh converging. As predicted by the fundamental kinematics, the parallel shear stress component vanishes at the moving contact line for quasi-stationary states (i.e.\ for $\dot\thetad=0$), and the dynamic contact angle is determined by a balance between the uncompensated Young stress and an effective contact line friction. Furthermore, a non-linear generalization of the model is proposed, which aims at reproducing the Molecular Kinetic Theory of Blake and Haynes for quasi-stationary states.
\end{abstract}


\section{Introduction}
\label{sec:intro}
\input{introduction.tex}

\section{Mathematical Modeling}
\label{sec:mathematical_model}
\input{modeling.tex}

\section{Numerical Methods}
\label{sec:Numeric_method}
\subsection{The Volume-of-Fluid method}
The Volume-of-Fluid (VOF) method for representing fluid interfaces coupled with a flow solver is well-known to be suited for solving interfacial flows (see e.g. \cite{Scardovelli1999,Popinet1999,Tryggvason2011,Maric2020}). We use the free software \href{http://basilisk.fr/}{\emph{Basilisk}}, a platform for the solution of partial differential equations on adaptive Cartesian meshes (\cite{Popinet2009,Popinet2015,Popinet2018}). 
For a two-phase flow, the volume fraction $c(\mathbf{x}, t)$ is defined as the integral of the first fluid's characteristic function in the control volume. The volume fraction $c(\mathbf{x}, t)$ is used to define the density and viscosity in the control volume according to
\begin{equation}\label{eq:basilisk_rho}
\begin{aligned} \rho({c}) & \equiv {c} \rho_1+(1-{c}) \rho_2, \\ \mu({c}) & \equiv {c} \mu_1+(1-{c}) \mu_2, \end{aligned}
\end{equation} 
with $\rho_1$, $\rho_2$ and $\mu_1$, $\mu_2$ the densities and viscosities of the phase 1 and 2 respectively.\\
The advection equation for the density is then replaced by the equation for the volume fraction
\begin{equation}\label{eq:basilisk_advection}
\partial_{t} c+ \mathbf{v} \cdot \mathbf{\nabla} c =0.
\end{equation}
The projection method is used to solve the incompressible Navier-Stokes equations combined with a Bell-Collela-Glaz advection scheme and a VOF method for interface tracking. The resolution of the surface tension term is directly dependent on the accuracy of the curvature calculation. The height-functions method, described in~\cite{Afkhami2008, Afkhami2009}, is a VOF-based technique for calculating interface normals and curvatures. About each interface cell, fluid ‘heights’ are calculated by summing fluid volume in the grid direction closest to the normal of the interface.
In two dimensions, a $7 \times 3$ stencil around an interface cell is constructed and the heights are evaluated by summing volume fractions horizontally, i.e.
\begin{equation}\label{eq:basilisk_heights}
h_{j}=\sum_{k=i-3}^{k=i+3} c_{j, k} \: \Delta,
\end{equation}
with $c_{j, k}$ the volume fraction and $\Delta$ the grid spacing. The heights are then used to compute the interface normal $\mathbf{n}_\Sigma$ and the curvature $\kappa$ according to
\begin{equation}\label{eq:basilisk_curvature}
\begin{array}{c}{\mathbf{n}_\Sigma=\left(h_{x},-1\right)}, \\  \\ {\kappa=\dfrac{h_{x x}}{\left(1+h_{x}^{2}\right)^{3 / 2}}}, \end{array}
\end{equation}
where $h_{x}$ and $h_{xx}$ are discretized using second-order central differences. The orientation of the interface, characterized by the contact angle -- the angle between the normal to the interface at the contact line and the normal to the solid boundary -- is imposed in the contact line cell. It is important to note that a numerical specification of the contact angle affects the overall flow calculation in two ways:
\begin{enumerate}
\item it defines the orientation of the interface reconstruction in cells that contain the contact line;
\item it influences the calculation of the surface tension term by affecting the curvature computed in cells at and near the contact line. 
\end{enumerate}
We now present the numerical implementation of the Generalized Navier Boundary Condition as written in \eqref{eqn:kinematics/regularized_gnbc}. The boundary condition is applied on the solid surface with a smoothing function that takes into account the relative position along the boundary with respect to the contact line, denoted $x$
\begin{align}
\label{eqn:basilisk_gnbc}
\beta (\mathbf{v}_\parallel-\vuwall) + (\mat{S}\ndomega)_\parallel + \dirac \: \sigma (\cos \thetad - \cos \thetaeq) \, \ngamma = 0 \quad \text{on} \quad \partial \Omega,
\end{align}
with $\dirac$ the \revgen{smoothed} Dirac function defined as
\begin{equation}\label{eq:GNBC_bell}
\dirac = \dfrac{\left( 1 - \tanh^2 \left( \dfrac{x}{\varepsilon} \right) \right)}{\varepsilon}.
\end{equation}
This specific function is smooth, symmetric and preserves the area for varying $\varepsilon$, characteristic features that are necessary for the well-posedness of the discrete boundary condition. The boundary condition can be expressed as an inhomogeneous Robin boundary condition for the parallel velocity $\mathbf{v}_\parallel$, as outlined above:
\begin{align}
\label{eqn:basilisk_gnbc2}
\mathbf{v}_\parallel + \dfrac{1}{\beta } (\mat{S}\ndomega)_\parallel = \vuwall + \dfrac{1}{\beta } \dirac \: \sigma (\cos \thetaeq - \cos \thetad) \, \ngamma \quad \text{on} \quad \partial \Omega.
\end{align}
We use the Navier boundary condition (Navier slip) that was implemented in the same framework in~\cite{Fullana2020} and tested as a localized slip boundary condition in~\cite{Lacis2020}. The difference lies now in the space dependent right-hand side of \eqref{eqn:basilisk_gnbc2}. The uncompensated Young's stress that only acts at the contact line through the discrete Dirac function, needs to be computed at each grid point.

\begin{figure}
\centering
\includegraphics[width=.8\textwidth]{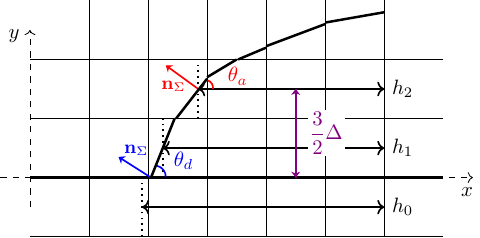}
\caption[Schematic of the angle extrapolation using the GNBC.]{Extrapolation of the contact angle $\thetad$ using the \revgen{angle above} $\theta_a$ located $3/2 \: \Delta$ away from the wall.$h_0$ to $h_2$ denote the horizontal heights.}
\label{fig:extrapolation}
\end{figure}

The numerical approach in this study stands out for its free contact angle method. Instead of setting the dynamic angle $\thetad$, we reconstruct it from the interface geometry and use it as an input parameter to calculate the right-hand side of \eqref{eqn:basilisk_gnbc2}. \revgen{To reconstruct such a consistent angle from the volume fraction field, we use a Taylor expansion of the contact angle along the coordinate direction normal to the boundary (see $y$-axis in figure~\ref{fig:extrapolation})
\begin{equation}\label{eq:taylor1}
    \theta_d = \theta_a + \delta \, \dfrac{\mathrm{d} \theta }{\mathrm{d} y} + O(\delta^2),
\end{equation}
where $\theta_a$ is the \revgen{angle above} located at the distance $y = \delta$ from the wall. Using that $\mathrm{d}\theta / \mathrm{d} s = \kappa$ and projecting onto the wall-normal direction, we obtain 
\begin{equation}\label{eq:taylor2}
    \theta_d = \theta_a + \delta \, \dfrac{\kappa \, \sqrt{1 + x^\prime}}{\sin \theta_a} \,  + O(\delta^2),
\end{equation}
with $x^\prime$ is the local slope of the interface. Discretely, by setting $\delta = 3/2 \, \Delta$, such that the \revgen{angle above} is computed one layer above the wall and using height-functions to compute the slope of the interface, such that $h_y$ is the first order derivative of the height-functions in the $y$ direction (normal to the wall), we obtain 
\begin{equation}\label{eq:angle_extrapolated}
\theta_{d} = \theta_{a} + \dfrac{3}{2} \Delta \dfrac{\kappa \: \sqrt{1+h_y^2}}{\sin\theta_{a}} + O(\Delta^2).
\end{equation}}
Figure~\ref{fig:extrapolation} provides a schematic illustration of this extrapolation process. Once the extrapolated angle is computed, we enforce it through appropriate local modification of height-functions in the ghost layer, similar to a regular contact angle. Algorithm~\ref{alg:angle_extrapolation} is a concise summary of the two-step procedure to apply the CR-GNBC in the VOF framework.

\begin{algorithm*}[ht!]
    \SetKwData{Left}{left}
	\SetKwData{This}{this}
	\SetKwData{Up}{up}
	\SetKwInOut{Input}{input}\SetKwInOut{Output}{output}

	\For{each boundary cell}{
    1. Locate the contact line cell\\
	2. Locate the cell one grid point above the contact line\\
	3. Compute the \revgen{angle above} $\theta_a$ using the unit normal $\mathbf{n}_\Sigma$\\
	4. Compute the first order derivative $h_x$ of the height-functions\\
	5. Compute the interface curvature $\kappa$\\
	6. Compute the extrapolated angle $\thetad$ using \eqref{eq:angle_extrapolated}\\
	}
    
	\BlankLine 7. Apply $\thetad$ at the contact line through height-functions\\

    \BlankLine
    
    \For{each boundary cell}{
    8. Compute the right-hand-side of \eqref{eqn:basilisk_gnbc2} using $\thetad$ and $\thetaeq$\\
	}

    \BlankLine 9. Apply the boundary condition for $\mathbf{v}_\parallel$ using \eqref{eqn:basilisk_gnbc2}\\
 
	\caption{CR-GNBC pseudo-code}\label{alg:angle_extrapolation}
\end{algorithm*}

\subsection{Kinematic transport of the contact angle}\label{subsection:validation-kinematics}

We validate the free contact angle method presented in \eqref{eq:angle_extrapolated} through an analysis of the kinematic transport of the contact angle in a simplified setup. Leveraging kinematic considerations,~\cite{Fricke2020} and \cite{Fricke2021} derived analytical solutions for the transport of the contact angle and the curvature for some specific velocity fields. To validate the present approach within the VOF framework, we conduct advection testcases for an initially circular interface in contact with the domain boundary. These advection tests are carried out for various grid sizes.

The setup involves a disk with a dimensionless diameter $D = 1$ in a $2 \times 2$ domain\secondrev{. The center of the disk is shifted by $\delta_s = 0.05$ above the substrate, resulting in an initial contact angle of $\theta_0 = \arccos(\delta_s/R) \approx 84.26^\degree$}. The velocity field across the entire domain is defined as:

\begin{equation}\label{eq:incompressible_velocities}
\begin{aligned}
v_x &= c_1 \cos(\pi t) \, x + c_2 \cos(\pi t) \, y, \\
v_y &= - c_1 \cos(\pi t) \, y.
\end{aligned}
\end{equation}

\noindent Here, $v_x$ and $v_y$ represent the $x$ and $y$ components of the velocity, while $c_1$ and $c_2$ are positive constants. We aim to validate the accuracy and reliability of the angle extrapolation method under varying grid sizes, where we only consider the advection equation of the color function \eqref{eq:basilisk_advection}.



\begin{figure}
\centering
\subfloat[]{
\centering
\includegraphics[width=0.47\linewidth]{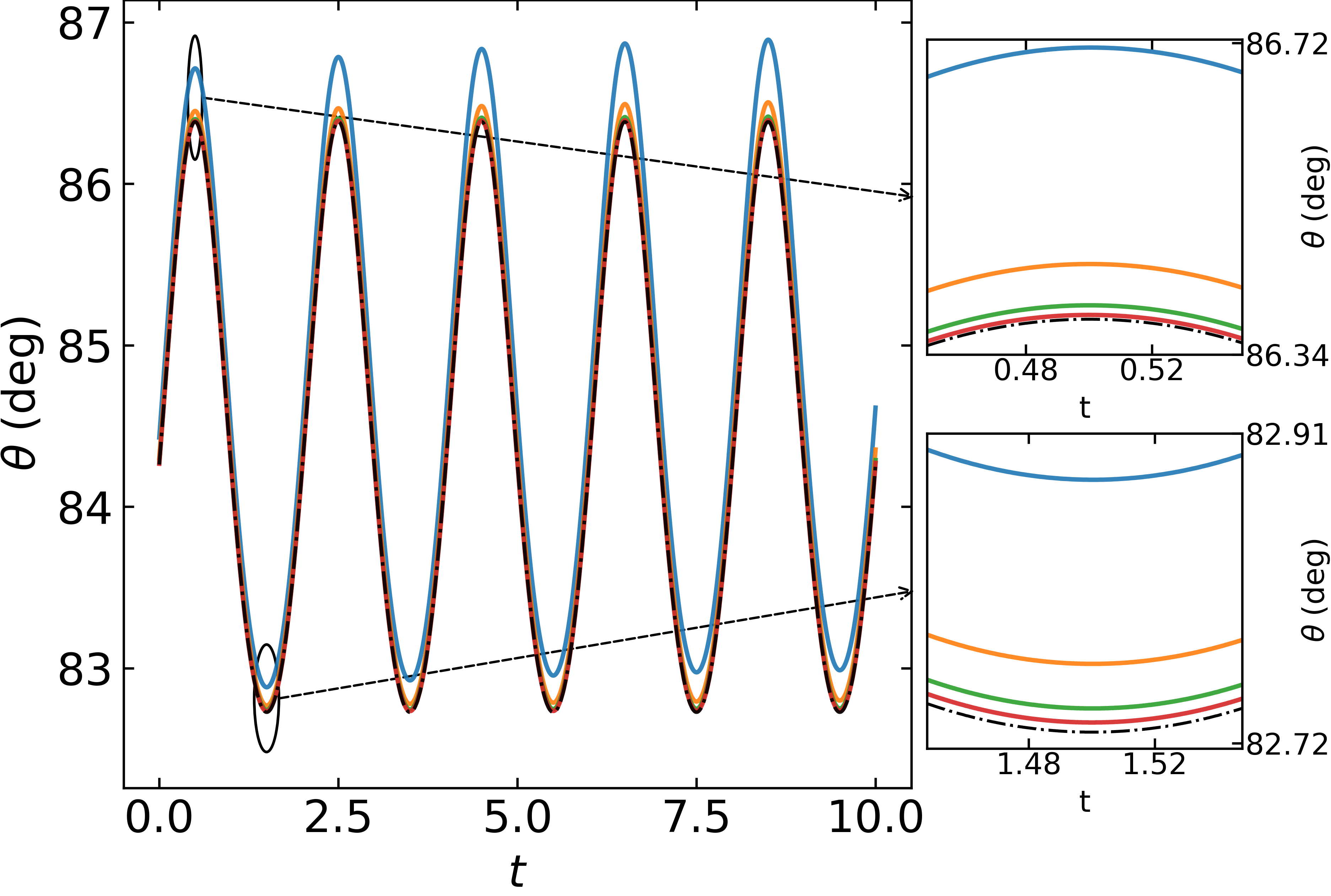}
}
\subfloat[]{
\centering
\includegraphics[width=0.47\linewidth]{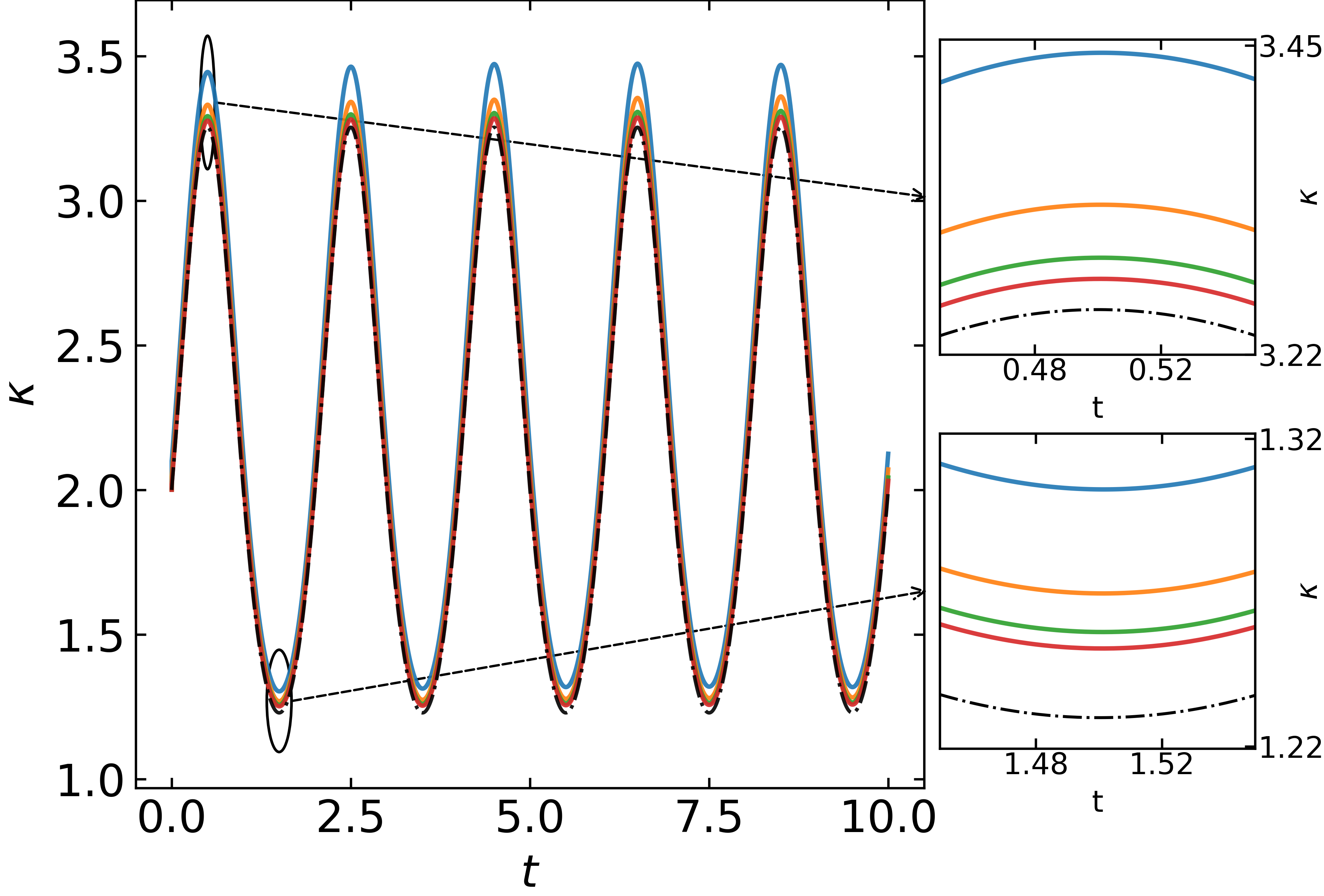}
}

\subfloat[]{
\centering
\includegraphics[width=0.6\linewidth]{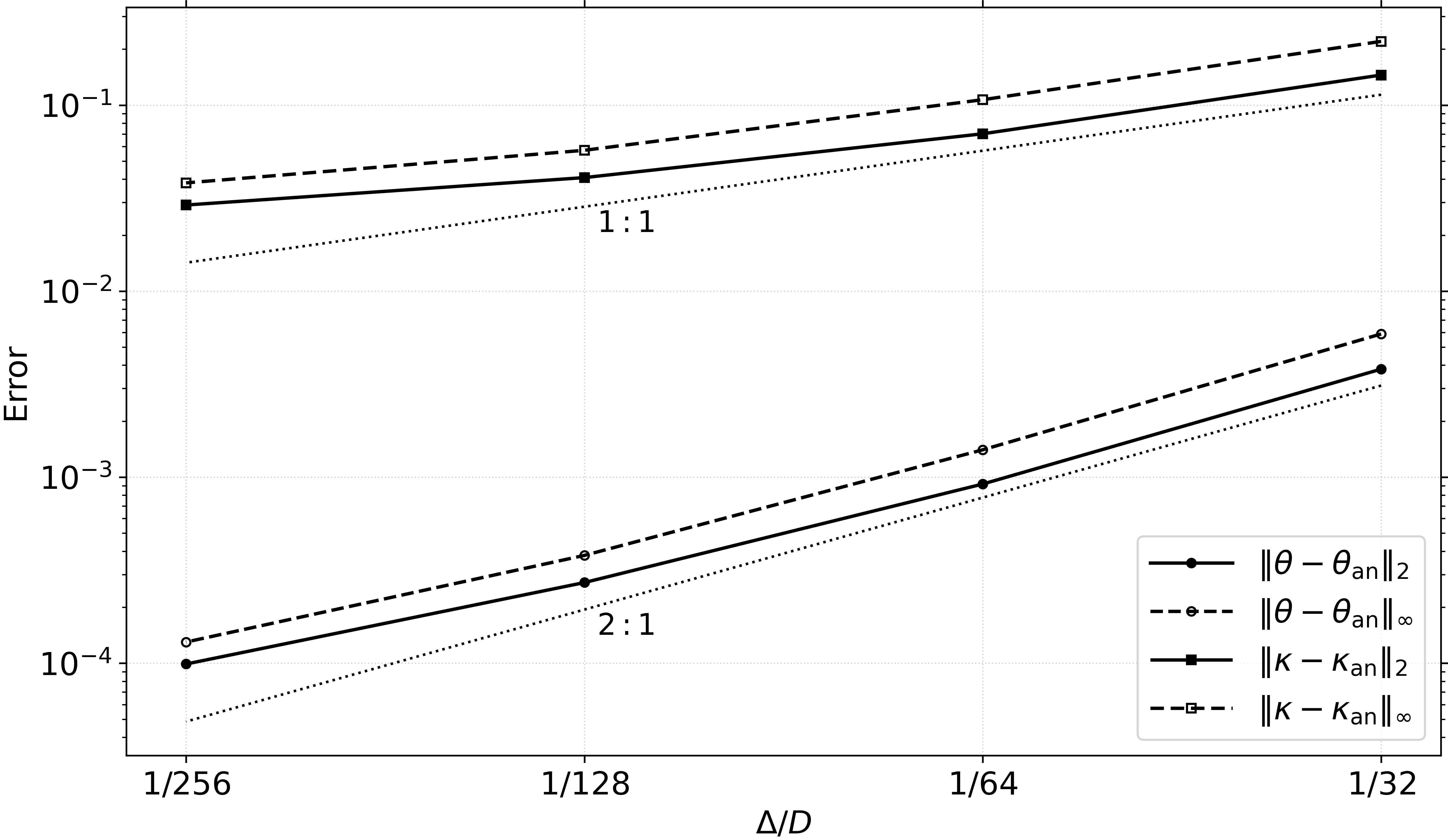}
}
    \caption{\secondrev{Validation of the free angle extrapolation method for varying grid sizes. (a) Temporal evolution of the contact angle. (b) Temporal evolution of the curvature.  In both panels, blue corresponds to $D/\Delta = 32$, orange to $D/\Delta = 64$, green to $D/\Delta = 128$, red to $D/\Delta = 256$, and the black dashed line is the analytical solution. (c) Convergence of both the angle and curvature errors with $L_2$ and $L_\infty$ norms, showing quasi-second-order convergence for the angle and quasi-first-order convergence for the curvature.}}
    \label{fig:angletransport}
\end{figure}

The prescribed incompressible velocity field \eqref{eq:incompressible_velocities} will induce oscillations of the interface in both vertical and horizontal directions. The angle formed at the contact line is determined by this motion and varies in time. From the relations derived in~\cite{Fricke2020}, we compare the observed numerical contact angle with the analytical value $\thetaan$, given by the formula
\begin{equation}\label{eq:angle_exact}
    \thetaan(t) = \dfrac{\pi}{2} + \tan^{-1} \left( \dfrac{- 1}{\tan \theta_0} e^{2 \, c_1 \, S(t)} + \dfrac{c_2}{2 \, c_1} \left( e^{2 \, c_2 \, S(t)} - 1 \right) \right)
\end{equation}
with
\begin{equation}\label{eq:S}
    S(t) = \dfrac{\sin (\pi t)}{\pi}.
\end{equation}
Moreover, we validate the evolution of the curvature by comparison with the reference one $\kappaan$, which is given as the solution of the ordinary differential equation (see \cite{Fricke2021})
\begin{equation}\label{eq:kappa_exact}
    \dfrac{d \kappaan}{d t} = -3 \, \kappaan \, \cos(\pi t) \, [c_1 \, \cos^2(\thetaan) - c_2 \, \cos(\thetaan) \sin(\thetaan) - c_1 \, \sin^2(\thetaan)]
\end{equation}
with the initial condition $\kappa_0 = 2 / D = 2$. 
\secondrev{We use Vofi (see ~\cite{BNA2016291}) to initialize the volume fraction field, which yields a more accurate initial curvature.}
\secondrev{We conduct simulations with $c_1 = 0.5$ and $c_2 = 0.2$, using the free angle extrapolation method. The simulations run until a final dimensionless time $T = 10$, and we examine the convergence of the method with grid sizes varying from 32 to 256 points per diameter (see figure~\ref{fig:angletransport}) with a fixed timestep $\delta t = 0.2  \, \Delta / \max(|c_1|, |c_2|)$. The extracted contact angles and curvatures show convergence towards the analytical solution. As shown in figure~\ref{fig:angletransport}(c), we observe quasi-second-order convergence for the contact angle and quasi-first-order convergence for the curvature, the latter is expected since $\kappa \sim \partial\theta/\partial s$, with $s$ the arc-length.}
\secondrev{We emphasize that this validation corresponds to a purely kinematic advection problem, in which convergence is expected to degrade by one order from interface position to contact angle and to curvature, since each quantity involves an additional spatial derivative of the interface geometry. Such behavior is specific to purely advective test cases, while in fully coupled dynamic simulations, surface-tension-induced regularization typically improves curvature convergence \citep{Popinet2009,Popinet2018}.}

\section{Results}
\label{sec:Results}
   \begin{figure}
   \centering
     \includegraphics[width=\textwidth]{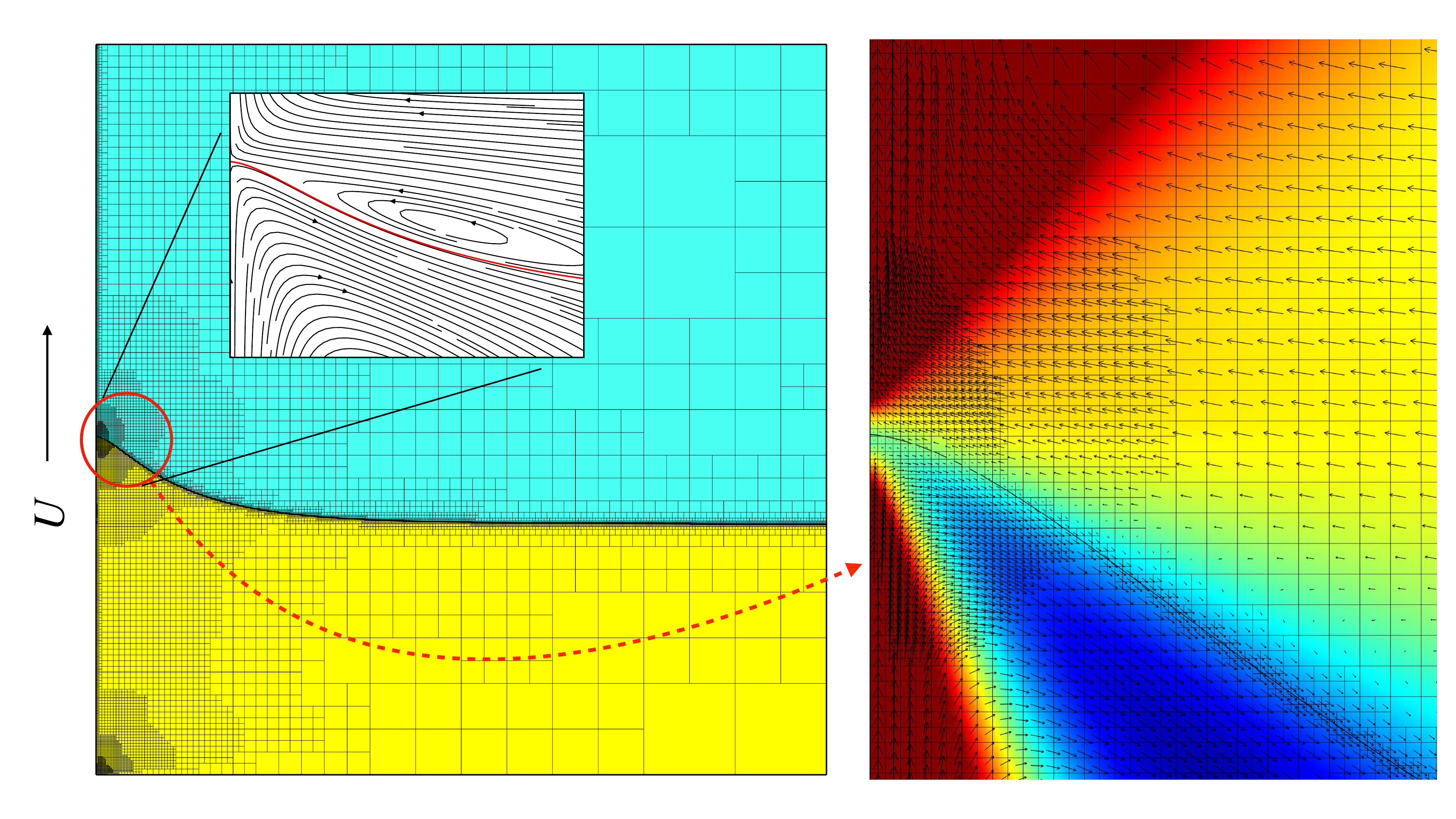}
        \caption{Steady-state meniscus example for $\cawall=0.1$ using the present CR-GNBC with $\varepsilon=0.05$. The image is in the contact line's reference frame, where the left plate is pulled up with \revgen{$\bar{\uwall} = \sqrt{\cawall}$}. The inset \revgen{shows a zoom around the contact line with streamlines highlighting a stagnation point in the upper phase}}
		\label{fig:CL_Flow_DNS_SAMPLE}
	\end{figure}
We apply the numerical method for the CR-GNBC model to the pulling plate setup, following the approach discussed in Section \ref{sec:intro}. This setup is akin to \revgen{setup B of}~\cite{Afkhami2018}. \revgen{We set the density ratio to $\rho_1/\rho_ 2 = 5$ and the viscosity ratio to $\mu_1/\mu_2 = 1$. The dimensionless capillary length is set to $\bar{l}_c = 1$, the dimensionless viscosities to $\bar{\mu}_1 = \bar{\mu}_2 = \bar{U}_w$ and the dimensionless density $\bar{\rho}_1 = 1$, such that the Reynols number is unity ($\operatorname{Re} = \bar{\rho}_1 \bar{l}_c \bar{U}_w / \bar{\mu}_1 = 1$) for any prescribed wall capillary number. In this configuration, setting the dimensionless surface tension to $\bar{\sigma} = 1$, the wall velocity is $\bar{U}_w = \sqrt{\cawall}$.}

Figure~\ref{fig:CL_Flow_DNS_SAMPLE} displays the results of a steady-state simulation. The image is presented in the reference frame of the contact line, where the contact line remains stationary while the left wall is pulled upwards. The velocity field relaxes, creating a stagnation point at the contact line. Additionally, the streamlines reveal another stagnation point formed above the interface in the lighter phase. It is worth noting that the characteristics of this additional stagnation point depend on the viscosity ratio, although our primary focus is not on this aspect.

\begin{figure}
\centering
    \includegraphics[width=0.9\textwidth]{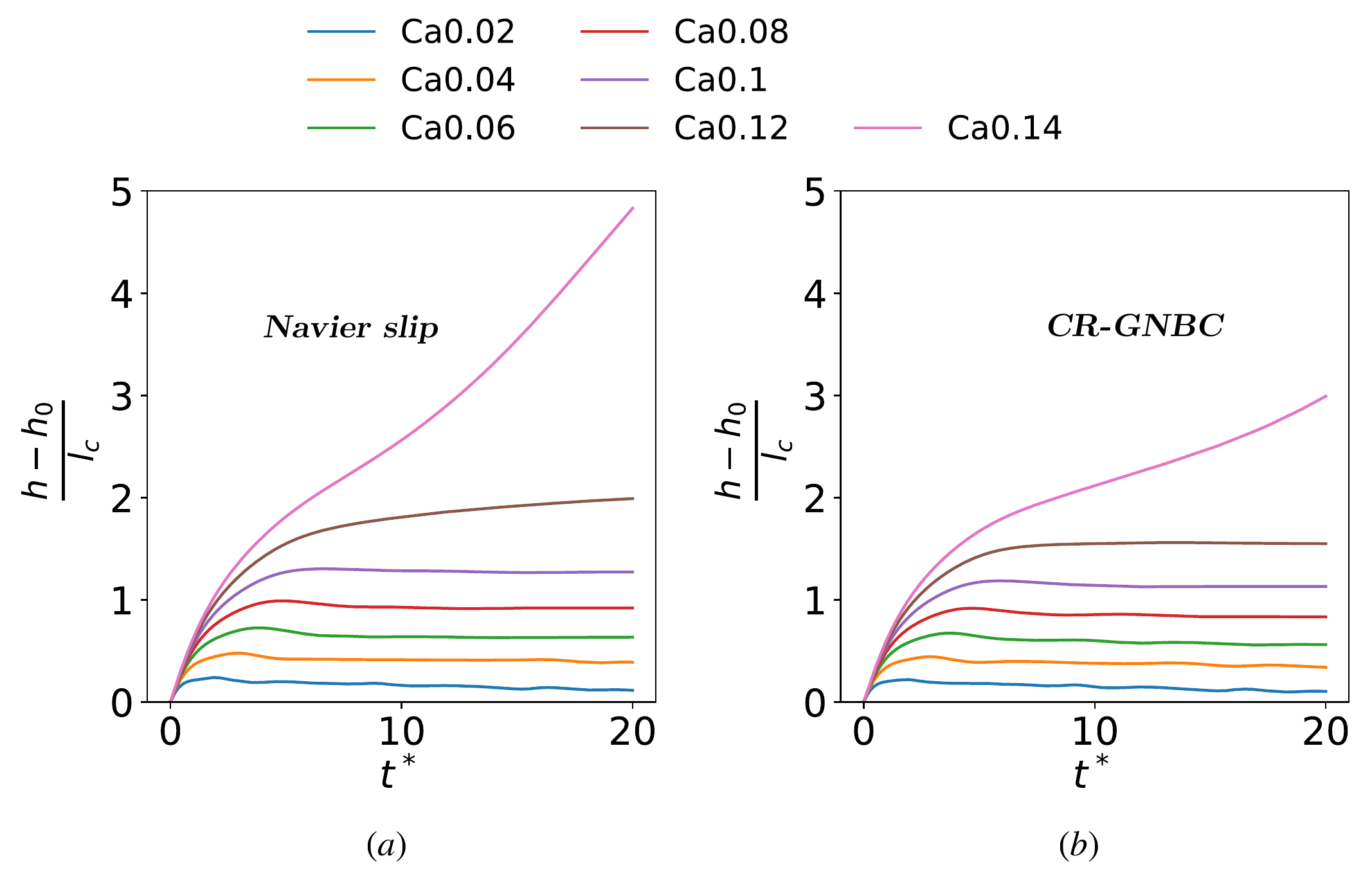}
    \caption{Vertical height of the contact line as a function of time for different capillary numbers $\cawall$, presented separately for (a) simple Navier boundary condition and (b) CR-GNBC. Steady-state heights are achieved, and a transition $\catr$ is observed, beyond which the liquid film rises continuously. In both (a) and (b), $\catr = 0.13$. Simulations are conducted with $\varepsilon = 0.05$, $\thetaeq = 90^{\degree}$, and a resolution of $\varepsilon / \Delta = 5.12$.}
    \label{fig:steady_state_heights}
\end{figure}

In the pulling plate setup, a distinctive characteristic is the presence of a de-wetting transition capillary number $\catr$, marking the point beyond which liquid film entrainment occurs, leading to an absence of a steady-state position for the contact line. Previous numerical results by \cite{Afkhami2018} identified this transition capillary number, but it was grid-dependent. Using the CR-GNBC method, with $\varepsilon$ resolved (i.e. larger than the grid size $\Delta$), we obtain a grid-independent $\catr$. This is depicted in figure~\ref{fig:steady_state_heights}, which shows the contact line position representing the fluid film height over time. For $\cawall \leq 0.12$, a steady-state height is eventually reached; however, for $\cawall=0.14$, the height continually increases. Thus, we determine that $\catr$ for this case is $\cawall=0.13 \pm 0.01$. The influence of Young's stress is evident when comparing figure~\ref{fig:steady_state_heights}(a) with figure~\ref{fig:steady_state_heights}(b). The $\catr$ remains the same, but the steady-state height exhibits a slight decrease. A convergence study demonstrating the grid independence of the CR-GNBC is presented in Appendix \ref{appex:convergence_study}. Furthermore, with the present approach, the parameters influencing $\catr$ are $\varepsilon$, the slip length $\lambda$ and the equilibrium contact angle $\thetaeq$. The dependence of these parameters on the $\catr$ is presented in Appendix \ref{appex:ca_tr_eps_GNBC}.

\subsection{Relaxation towards steady state}

\begin{figure}
\centering
    \subfloat[]{
    \centering
    \includegraphics[width=0.45\linewidth]{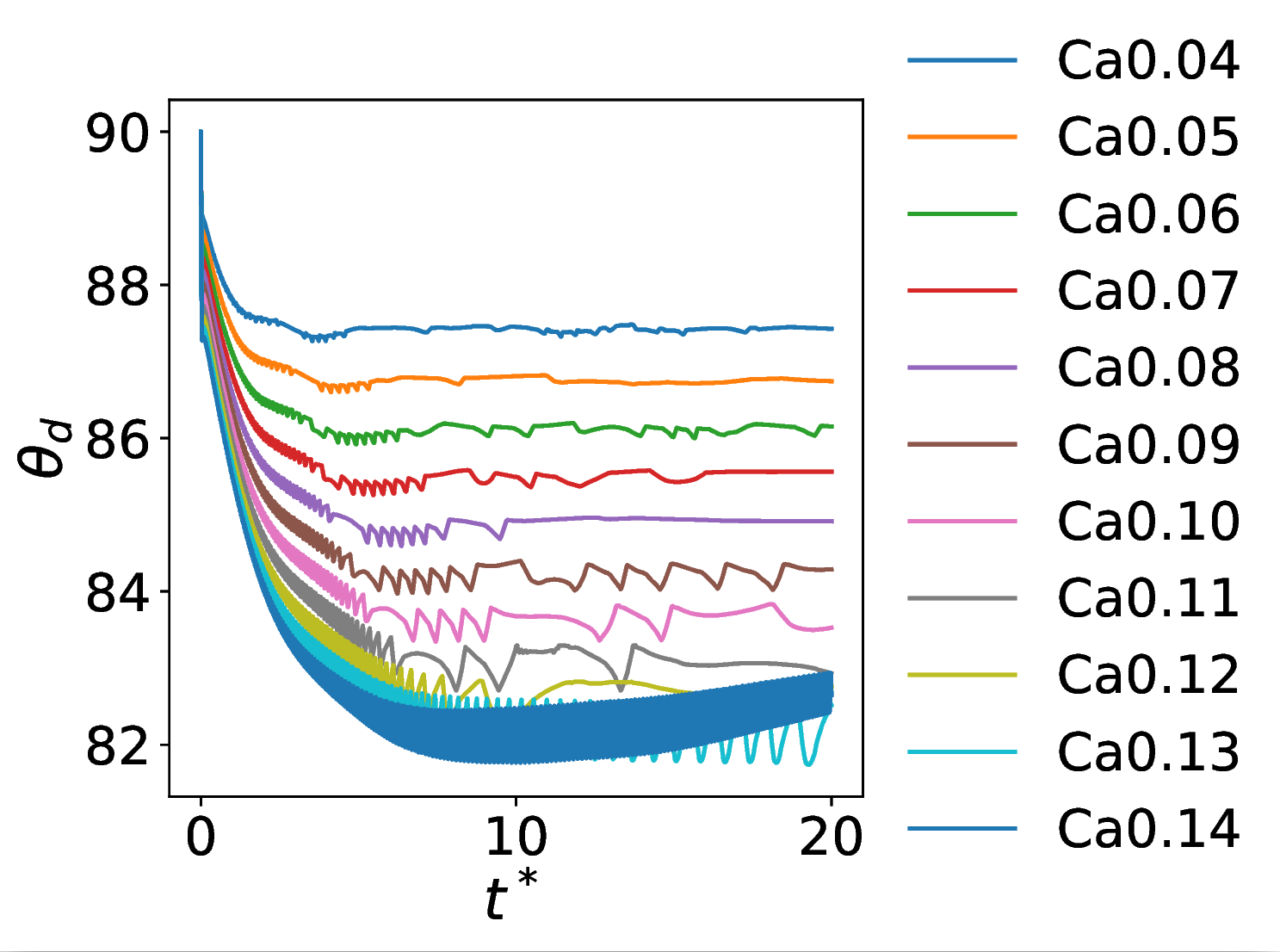}
    }
    \subfloat[]{
    \centering
    \includegraphics[width=0.45\linewidth]{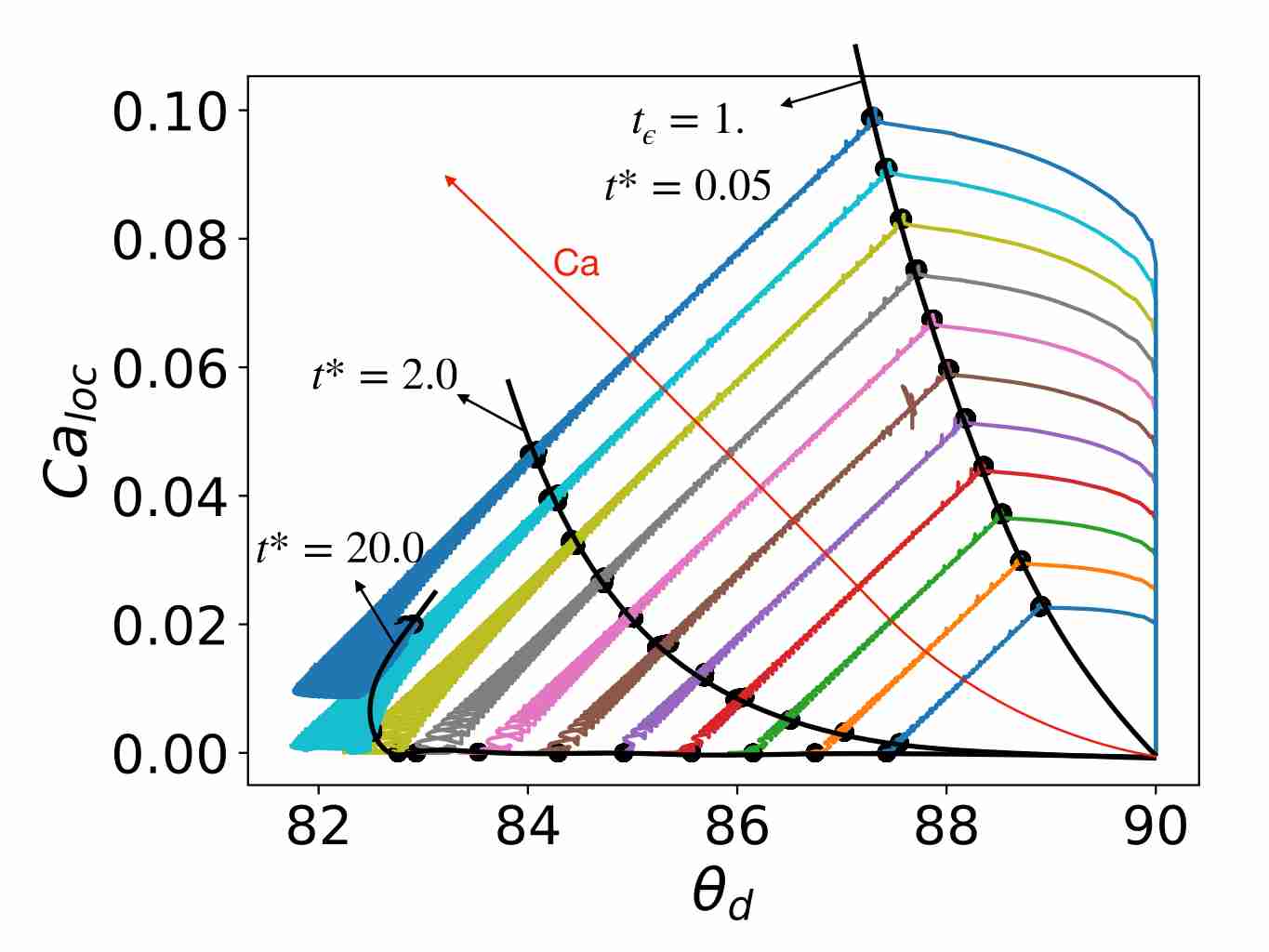}
    }
    \caption{(a) Evolution of the dynamic contact angle $\thetad$ in the CR-GNBC simulation for various $\cawall$. The angle begins to deviate from the initial value of $90^\degree$ and eventually reaches a steady state. Around $\catr$, the angle exhibits oscillations over time. (b) The relaxation plot on a $\theta-\cawall$ plane. Here, $\caloc$ represents the contact line capillary number. Time progresses from right to left, and a maximum in $\caloc$ is reached at $t_{\varepsilon} = 1$, which corresponds to the slip length timescale ($\varepsilon / \uwall$). After this point, $\caloc$ starts relaxing towards a steady state ($\caloc = 0$). Above $\catr$, $\caloc$ reaches a minimum and starts rising again. This set of simulations are the same as in figure~\ref{fig:steady_state_heights}(b).}
    \label{fig:FULL_GNBC_angle_relaxation_Ca_loc}
\end{figure}

Starting from a horizontal two-fluid interface at rest, we now compare the transient characteristics. A distinctive feature of the present CR-GNBC model is that the contact angle is not fixed \emph{a priori}. Figure~\ref{fig:FULL_GNBC_angle_relaxation_Ca_loc}(a) illustrates the contact angle $\thetad$ as a function of time. The angle initiates at $90^\degree$ and subsequently relaxes to a steady-state value different from $90^\degree$. Despite converging to a steady state, the observed value of $\thetad$ exhibits spurious oscillations. These oscillations intensify with increasing $\cawall$; however, their influence is minor, with amplitudes remaining below $0.5^\degree$ and diminishing with grid refinement.
In figure\ref{fig:FULL_GNBC_angle_relaxation_Ca_loc}(a), we observe an interesting trend when plotting $\thetad$ against $\caloc$, as shown in figure~\ref{fig:FULL_GNBC_angle_relaxation_Ca_loc}(b). Here, $\caloc$ represents the contact line $Ca$ in the lab frame. It starts at $0$ since everything is initially at rest and eventually returns to $0$ in a quasi-stationary state. During the transient phase, although we set the solid velocity to $\uwall$ instantly, $\caloc$ takes some time to reach its maximum value. This time, defined as $t_{\varepsilon} = \varepsilon/\uwall$, represents a relaxation timescale due to contact line friction. While a detailed examination of the behavior for $t < t_{\varepsilon}$ is beyond the current study's scope, we observe that, once $\caloc$ reaches its peak, it begins to relax to the steady state where $\caloc = 0$, and $\thetad$ follows the GNBC law \eqref{eqn:kinematics/empirical_relation_gnbc_v1}.

\begin{figure}
    \centering
    \includegraphics[width=0.85\textwidth]{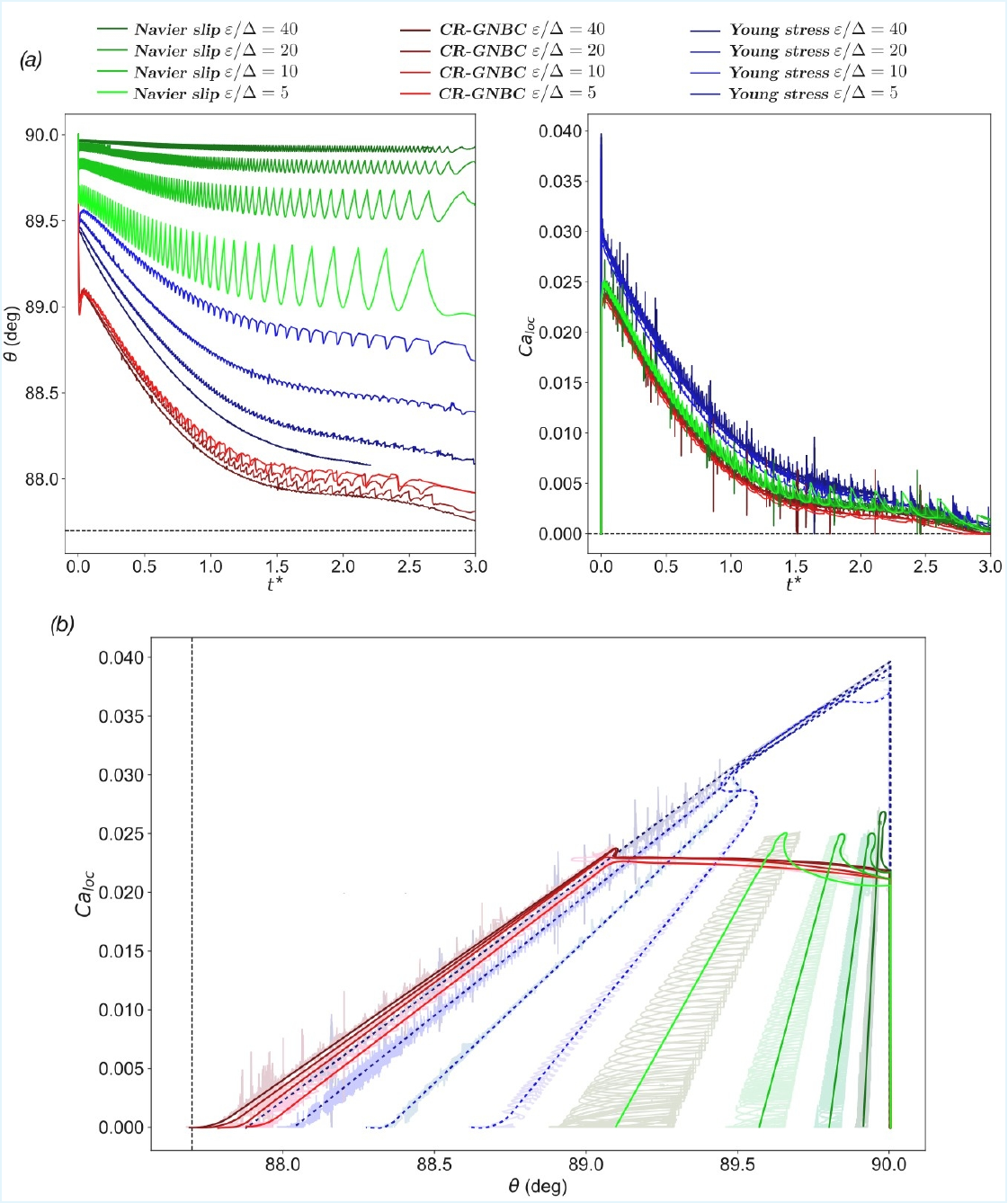}
    \caption{The relaxation plots Navier slip (green curves), CR-GNBC (red curves) and no-slip with Young stress (blue curves). All the plots are done for $\cawall=0.04$ and $\varepsilon=0.05$. The grid resolution is reported in terms of $\varepsilon / \Delta$, and color intensity is increased to show higher resolution. Figure (a) shows the contact angle $\thetad$ and the contact line speed in the lab frame of reference $\caloc$ as a function of time. In (b) we show the phase diagram resulting from figure (a). The dashed black line represents the GNBC law angle in the steady state. Time flows from right to left and aligns the curves. Each curve set has its own characteristic feature. The oscillations, present in (a), are faded in the phase diagram for clarity.}
\label{fig:ALL_GNBC_SLIP_NO_SLIP_angle_relaxation}
\end{figure}

In figure~\ref{fig:ALL_GNBC_SLIP_NO_SLIP_angle_relaxation}, we illustrate the behavior of each term of the CR-GNBC equation \eqref{eqn:basilisk_gnbc2}. We analyze and present each outcome for three different boundary conditions:

\begin{enumerate}
    \item Navier slip with a constant contact angle $\thetad = \thetaeq$: 
        \begin{align}
        \label{eqn:GNBC_parts_NBC}
        \mathbf{v}_\parallel + \dfrac{1}{\beta } (\mat{S}\ndomega)_\parallel = \vuwall \quad \text{on} \quad \partial \Omega, 
        \end{align}
    \item No slip with uncompensated Young stress, with the "free angle" method~\eqref{eq:angle_extrapolated}:
        \begin{align}
        \label{eqn:GNBC_parts_YS_ONLY}
        \mathbf{v}_\parallel  = \vuwall + \dfrac{1}{\beta } \dirac \: \sigma (\cos \thetaeq - \cos \thetad) \, \ngamma \quad \text{on} \quad \partial \Omega,
        \end{align}

    \item Full CR-GNBC as written in~\eqref{eqn:basilisk_gnbc2} which combines contributions from both the above cases.
\end{enumerate}

We conducted simulations for each individual case (i), (ii), and (iii) and illustrate the behavior of each term in figure~\ref{fig:ALL_GNBC_SLIP_NO_SLIP_angle_relaxation}.
In figure~\ref{fig:ALL_GNBC_SLIP_NO_SLIP_angle_relaxation}(a) we see the angle as a function of time. Since we start from a horizontal surface, all plots begin at $90^\degree$. The green curves, representing the Navier slip case, converge to the constant imposed value of $90^\degree$. The relaxation to a steady-state angle is accompanied by oscillations, whose amplitude decreases with grid refinement. The blue curves, representing the behavior of uncompensated Young's stress with a no-slip boundary condition, show the effect of the free contact angle. Because the Young stress term involves the free contact angle method, the steady-state angle differs from $90^\degree$ and relaxes to the GNBC law contact angle as the grid is refined. These spurious oscillations are less pronounced than in the Navier slip case. Finally, the red curves represent the full CR-GNBC model. At the same level of grid refinement, the CR-GNBC model outperforms the Young stress case (blue curves) by being closer to the expected GNBC law contact angle and outperforms the Navier slip (green curves) by having fewer spurious oscillations.
In the plot of $\caloc$ vs time, we see that, although all the curves eventually relax to the steady state of $\caloc = 0$, there is a difference in the initial relaxing stage. As soon as the simulation is started we see that since blue curves have no slip, they rise to the $\caloc=\cawall$ in dimensionless time $t_{\varepsilon}$ and then relax to the steady state value, while the CR-GNBC and slip cases rise to the value equal to $\caloc <\cawall$.

Figure \ref{fig:ALL_GNBC_SLIP_NO_SLIP_angle_relaxation}(b) shows the phase diagram on a $\caloc-\theta$ plane. This figure sums up the the overall behaviour of the contact line dynamics in each case and a characteristic behaviour of each set could now be identified. The timeline in this figure progresses from right to left. 

\begin{enumerate}
    \item In the Navier slip case, we observe that at $t = 0$ and for $\thetad = 90^\degree$, when the interface is horizontal, $\caloc$ is null. Then, $\caloc$ suddenly rises to a maximum value, which remains lower than the imposed $\cawall$. This rapid rise occurs within the relaxation time $t_{\varepsilon}$, where $\varepsilon$ is the slip length. This behavior aligns with the discussion in figure~\ref{fig:FULL_GNBC_angle_relaxation_Ca_loc}(b). Subsequently, the contact line relaxes to a steady state where $\caloc$ returns to zero. This relaxation is accompanied by spurious oscillations in the contact angle $\thetad$. Ideally, in this case, the system should relax to $\thetad = 90^\degree$ throughout the motion and also in the steady state (given that we impose a constant $\thetad = \thetaeq= 90^\degree$), which is indeed observed as the grid is refined. The final angle $\thetad$ converges  to $90^\degree$, and spurious oscillations diminish with increasing grid refinement.
    \item In the no-slip with Young's stress, we notice an interesting pattern. At the start (t=0), the simulation begins with $\caloc = 0$ and $\thetad = 90^\degree$ at the lower right of figure~\ref{fig:ALL_GNBC_SLIP_NO_SLIP_angle_relaxation}. However, as soon as we advance in time, $\caloc$ increases to a maximum value equal to $\cawall$, and then, starts relaxing to $0$. With the presence of uncompensated Young stress, it ideally should relax to the GNBC law contact angle indicated by the dashed line in figure~\ref{fig:ALL_GNBC_SLIP_NO_SLIP_angle_relaxation}. We observe that oscillations are decreasing with grid refinement and the final value of the contact angle is converging towards the GNBC law angle.
    \item In the CR-GNBC case, we observe characteristics from both (i) and (ii). Initially, both $\caloc$ and $\thetad$ start from zero. Subsequently, $\caloc$ reaches a maximum during the relaxation time and eventually relaxes to the GNBC law contact angle. The notable advantage of the CR-GNBC is that even with a modest resolution of 5 grid points per slip length, the spurious oscillations, compared to case (i) at the same resolution, are significantly reduced. Moreover, the accuracy in relaxing towards the GNBC law contact angle (dashed line) is substantially improved compared to case (ii). Further grid refinement leads to a continued reduction in spurious oscillations and enhances accuracy.
\end{enumerate}

\subsection{Steady-state contact line dynamics: the GNBC smoothing signature}

\begin{figure}
\centering
    \includegraphics[width=\textwidth]{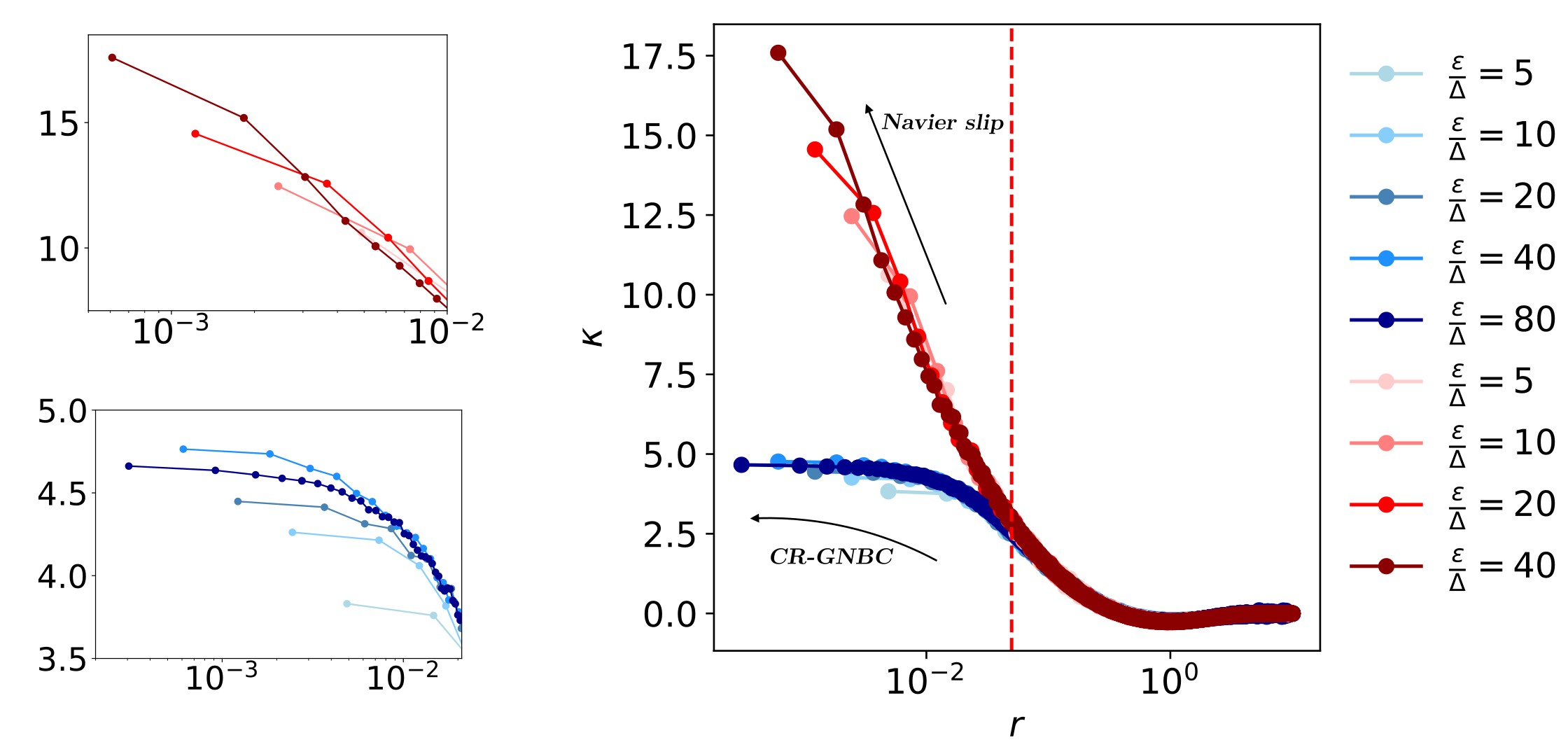}
    \caption{Curvature profiles relative to the radial distance from the contact line. The red curves represent curvature under the Navier slip boundary condition (slip), showing a logarithmic divergence. In contrast, the blue curves (GNBC) demonstrate the convergence to a finite curvature value and, thus, the removal the singularity present in the NBC. Simulations are conducted with $\cawall=0.08$ and $\varepsilon=0.05$. The equilibrium angle is $\thetaeq= 90^\degree$, and $\Delta$ denotes the grid size. Various color intensities denote grid refinement, where lighter shades correspond to a coarse mesh, and darker shades indicate a fine mesh.}
\label{fig:curvature_singularity_SLIP_GNBC}
\end{figure}

We now demonstrate the full regularization of the contact line singularity achieved by the present CR-GNBC method. Figure~\ref{fig:curvature_singularity_SLIP_GNBC} presents the curvature as a function of the distance from the contact line for various grid resolutions. The Navier slip model exhibits a logarithmic divergence in curvature, consistent with the analytical findings of \citet{devauchelle_josserand_zaleski_2007} and \citet{Kulkarni2023}. While the singularity in the Navier slip model is integrable and considered ``weak'', it induces a pressure singularity, rendering the slip model physically ill-posed. In contrast, the present CR-GNBC model regularizes the logarithmically singular curvature at the contact line ($\kappa \sim \log r$), establishing it as a physically well-posed model.

\begin{figure}
\centering
    \includegraphics[width=\textwidth]{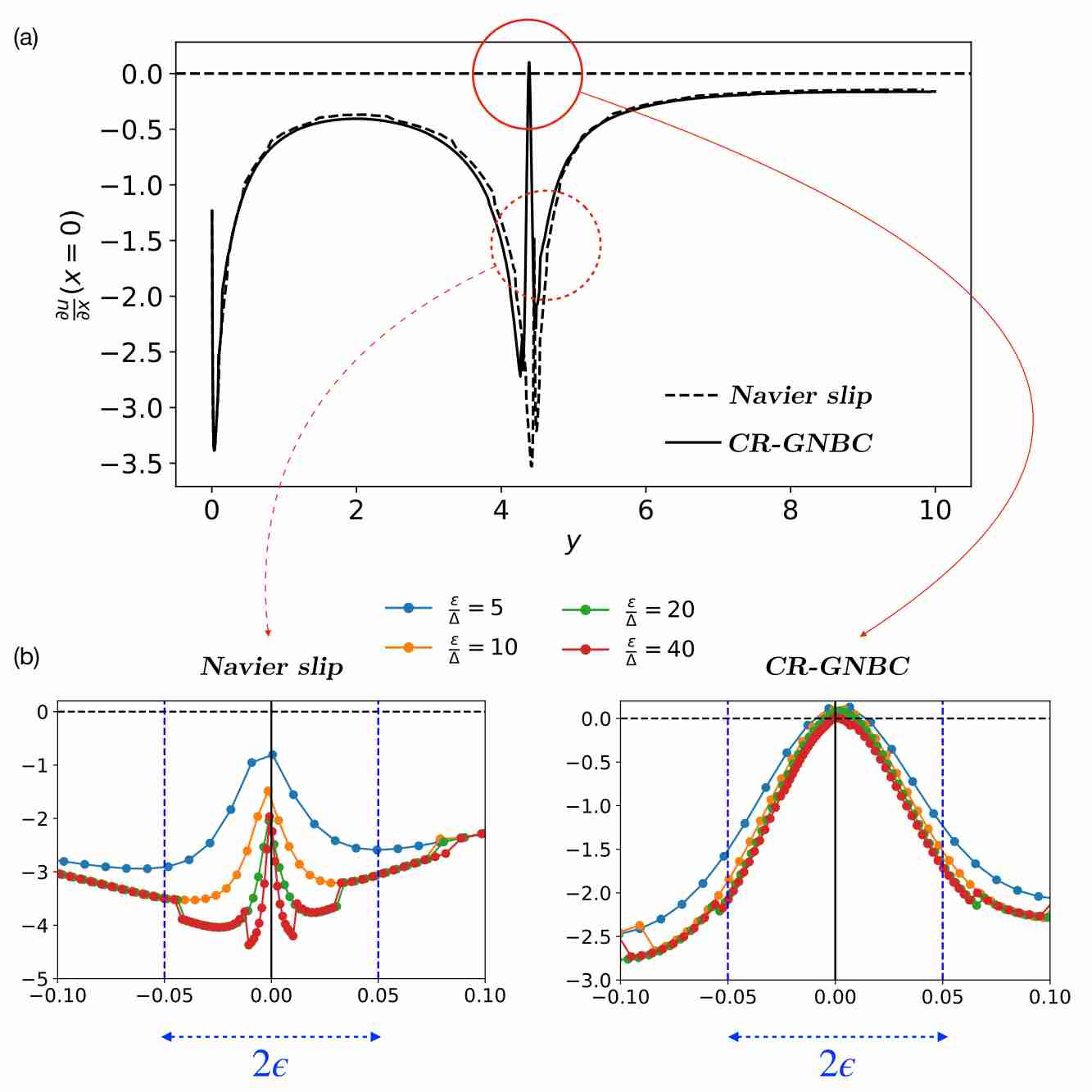}          
    \caption{(a) Wall shear stress in a steady-state simulation plotted against the vertical position $y$. The dashed line corresponds to the Navier slip boundary condition, while the solid line corresponds to the CR-GNBC case. Both curves largely overlap, except for a small region shown in the zoom-ins for Navier slip and CR-GNBC in (b). The zoomed-in figures are normalized by the contact line position, where $0$ on the $x$-axis corresponds to the contact line position. Notably, in the CR-GNBC case, the shear stress at the contact line is zero, whereas it is not the case for the Navier slip. The simulations are conducted with fixed $\cawall=0.08$ and $\varepsilon=0.05$ and varying grid sizes.}
\label{fig:shear_stress_GNBC_SLIP_compare}
\end{figure}
In Section~\ref{sec:mathematical_model}, we showed that assuming a $C^1$ velocity field up to the contact line in the reference frame of the moving wall, the rate of change of the contact angle scales with the shear stress at the contact line. In steady state, where $\dthetad = 0$, the shear stress must approach zero as it reaches the contact line. A non-zero shear stress would indicate a violation of the smoothness assumption made by \cite{Fricke2019}. This violation occurs in the Navier slip model, as non-zero shear stress is necessary for contact line motion. In figure~\ref{fig:shear_stress_GNBC_SLIP_compare}, we observe the behavior of shear stress for both Navier slip and CR-GNBC in steady state. At the contact line, the shear stress converges to zero within the $\varepsilon$ region in the CR-GNBC case, aligning with the expected smoothness of the flow field. However, for the Navier slip model, the shear stress fails to converge to zero.

 \begin{figure}
        \centering
        \includegraphics[width=0.7\textwidth]{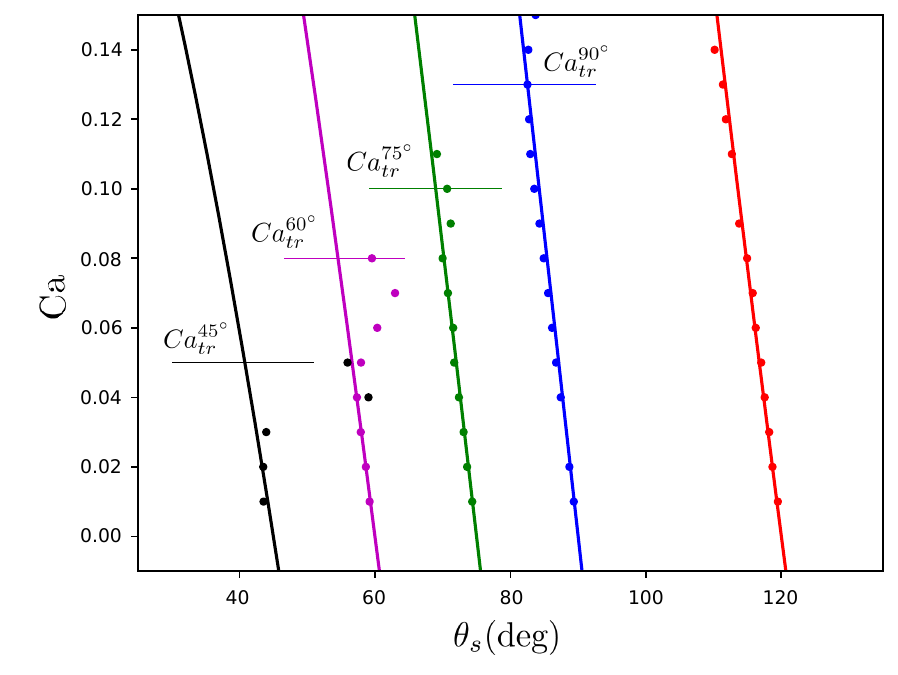}        \caption{The behavior of the quasi-stationary value of $\thetad$ vs. $\cawall$ is illustrated for various $\thetaeq$ and compared with the GNBC law \eqref{eqn:kinematics/empirical_relation_gnbc_v1}. The solid lines represent the analytical expression of the steady-state behavior expected from \eqref{eqn:kinematics/empirical_relation_gnbc_v1}, while the dots depict simulation results. The different colors represent various $\thetaeq$, progressing from left to right (black to red) as $45^\degree$, $60^\degree$, $75^\degree$, $90^\degree$, and $120^\degree$, respectively. The horizontal lines denote the $\catr$ for each equilibrium angle considered. An excellent agreement between simulations and the GNBC law is observed up to $Ca<\catr$.}
        \label{fig:GNBC_law}
    \end{figure}

Having demonstrated that the shear stress at the contact line in the steady state is zero using the CR-GNBC, we proceed to compare the quasi-stationary state GNBC relation \eqref{eqn:kinematics/empirical_relation_gnbc_v1} with our simulation results in figure~\ref{fig:GNBC_law}. Remarkably, we observe excellent agreement between the simulation outcomes and the quasi-stationary GNBC law, particularly for $\cawall < \catr$. It is essential to note that the behavior of $\cacl = f(\thetasteady)$ in figure~\ref{fig:GNBC_law}, as predicted by the quasi-stationary GNBC law \eqref{eqn:kinematics/regularized_gnbc_v2}, is not explicitly imposed but is a direct outcome from the simulations.

\section{Conclusion}
\input{conclusion.tex}

\section{Outlook}
\subsection{A non-linear generalization of the GNBC}\label{sec:non-linear-gnbc}
\input{outlook-nonlinear-gnbc.tex}

\begin{figure}
    \centering
    \includegraphics[width=0.5\textwidth]{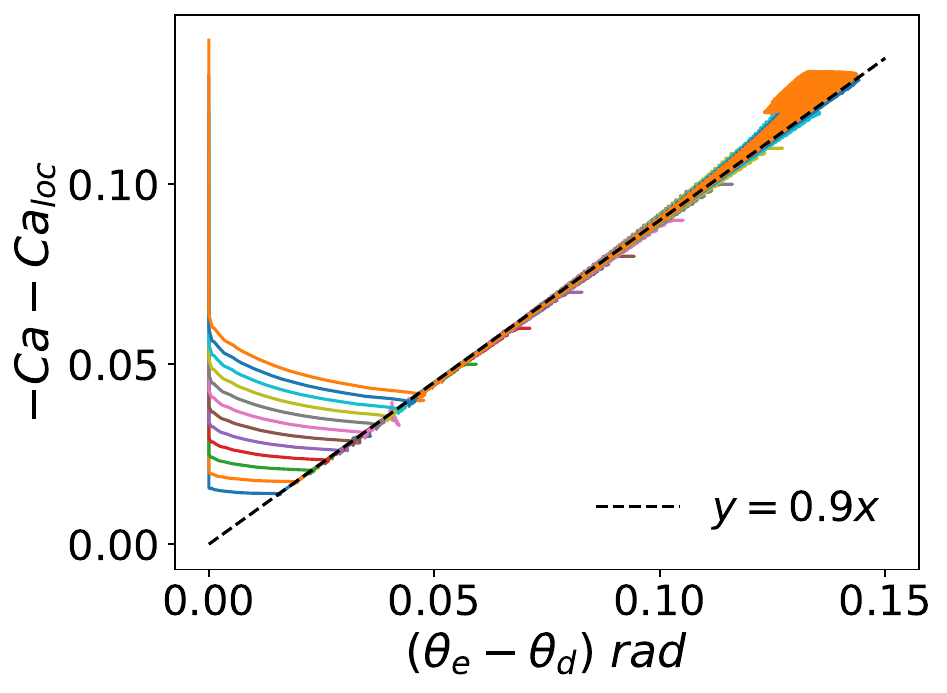}
    \caption{\revgen{Collapse of shifted $\caloc$ as a function of the angle deviation $\thetaeq - \thetad$}}
    \label{fig:Ca_loc_shift}
\end{figure}

\revgen{
\subsection{Relaxation time scale of the GNBC}
We can see from figure \ref{fig:FULL_GNBC_angle_relaxation_Ca_loc}b that after $\caloc$ reaches a maximum value within a short-lived impulsive slip length time-scale $t_{\varepsilon}$, there is an intermediate process of capillary relaxation before entering the slow decaying relaxation to the steady state. We refer to the timescale governing this intermediate relaxation as the GNBC relaxation time scale $t_r$. This timescale is crucial as it controls how fast a system will relax to the quasi-stationary state locally. Hence, we now examine what determines this time scale. Our formulation of the CR-GNBC model gives a contact angle evolution law \ref{eqn:kinematics/theta_evolution_gnbc} which can be re-written in terms of capillary number as 
\begin{equation}
\frac{\mu}{\sigma} \dthetad = - \frac{\cacl}{2 \, \lambda} + \frac{1}{2 \, \varepsilon} \,  (\cos \thetad - \cos \thetaeq),
    \label{eqc:angle_evolution_Ca}
\end{equation}
with $\cacl$ the contact line capillary number defined in the reference frame of the solid. In the lab frame of reference we can use $\caloc$ and $\cawall$ to write the angle evolution as 
\begin{equation}
  \frac{\mu}{\sigma}  \dthetad =   \frac{\cawall - \caloc}{2 \, \lambda} + \frac{1}{2 \, \varepsilon} \,  (\cos \thetad - \cos \thetaeq).
    \label{eqc:angle_evolution_Ca_Ca_loc}
\end{equation}
Assuming small deviations of $\theta_d$ from $\theta_e$, we get
\begin{equation}
\frac{\mu}{\sigma} \dthetad = \frac{\cawall - \caloc}{2 \, \lambda} + \frac{\sin \thetaeq}{2 \, \varepsilon} (\thetaeq - \thetad).
\label{eqc:angle_evolution_small_deviation}
\end{equation}
Note that in above equation, the time dependent quantities as $\thetad$ , $\dthetad$ and $\caloc$ require the solution of the Stokes flow equation subject to this boundary condition which leaves us only with numerical tools. However, motivated by the results of our DNS, we propose an asymptotic expansion of the contact line capillary number in terms of the deviation of the contact angle away from the equilibrium angle. From this expansion, we can also see that the contact line is mainly driven by the uncompensated Young stress in this relaxation time scale.
In the relaxation regime one may show that the contact‐line velocity admits a regular expansion in the small angle deviation
\begin{equation}
\delta = \thetaeq - \thetad\,,
\end{equation}
of the form
\begin{equation}
    \mu \, U_{cl} = A_0 (\sigma \delta) + A_1  ({\sigma \delta})^2 + A_2  ({\sigma \delta})^3 + \mathcal{O}(\delta^4),
\label{eqc:mobility_U_cl_small_deviation}
\end{equation} where $\mu$ is the dynamic viscosity, $\sigma$ the surface tension, and the coefficients $A_{i}$ are constants.
This expansion in \eqref{eqc:mobility_U_cl_small_deviation} applies strictly on the relaxation timescale $t_r$, over which the local capillary number $\caloc$ monotonically decreases from an initial value toward zero. Prior to this relaxation, an impulsive timescale of order $t_{\varepsilon}$ governs the rapid rise of $\caloc$ from zero to a peak value $\cawall_M$, accompanied by a corresponding change of the dynamic contact angle from $\thetaeq$ to some $\theta_{M}$.
It is important to note that \eqref{eqc:mobility_U_cl_small_deviation} is derived in the frame in which the solid substrate is stationary.  In the lab frame, where the plate translates at a constant capillary number $\cawall$, the steady–state contact‐line condition requires a finite offset in the boundary condition. The mobility law at the first order in relaxation time scale thus becomes 
\begin{equation}
    -\caloc = A_0 (\thetaeq - \thetad) + \Phi(\cawall),
    \label{eq:Ca_loc_offset}
\end{equation}
where $\Phi(Ca)$ is the offset determined by the system's steady-state behaviour. Substituting the above expansion in \eqref{eqc:angle_evolution_small_deviation}, we obtain
\begin{equation}
    \frac{\mu}{\sigma} \dthetad = \frac{ A_0 (\thetaeq - \thetad) }{2 \, \lambda} + \frac{\sin \thetaeq}{2 \, \varepsilon} (\thetaeq - \thetad)  + f(\cawall),
    \label{eqc:ode_theta_d_theta_e}
\end{equation}
where the steady-state behaviour, or GNBC law \eqref{eqn:kinematics/empirical_relation_gnbc_v1}, gives $f(\cawall) = - \dfrac{\varepsilon}{\lambda} \dfrac{\cawall}{\sin \thetaeq}$. The solution of the above ordinary differential equation is,
\begin{equation}
    (\thetaeq - \thetad) = C_{0} \, e^{-\omega t}  - \frac{\varepsilon}{\lambda} \frac{\cawall}{\sin \thetaeq},
\label{eqc:sol_ode}
\end{equation}
where
\begin{equation}
\omega = \dfrac{\sigma}{2 \mu} \left( \dfrac{A_0}{\lambda} + \dfrac{\sin \thetaeq}{\varepsilon}\right) .
    \label{eqn:omega_relaxation_time}
\end{equation}
The solution tells us that the relaxation time constant $t_r = 1/\omega $ has a contribution from both the viscous stress $\sim \sigma A_0 / \mu \lambda$ and Young stress $\sim \sigma \sin \thetaeq / \mu \varepsilon$. Since $A_0$ and $\sin \thetaeq$ both are of order one, the time scale is of the order of slip length-based capillary number and a contact region-based capillary number that comes out to be a few milliseconds for our flow parameters. 
The constant $C_0$, a function of $\cawall$, needs to be obtained by asymptotic matching to the impulsive time response. We leave the solution at the impulsive time scale $t_{\varepsilon}$ and a matched asymptotic solution at the full range of scales as future scope.\\
We now verify the existence of this relaxation timescale $t_r$ in our DNS. Re-plotting figure \ref{fig:FULL_GNBC_angle_relaxation_Ca_loc} with shifts accounting for each $\cawall$, we see that all the curves collapse. Hence, figure \ref{fig:Ca_loc_shift} verifies the region's existence, which happens at the relaxation time scale, where the linear dependence in \eqref{eq:Ca_loc_offset} is justified. We get $A_0 = 0.9$ from the best fit. 
At $t = 0$ for the relaxation time scale, we have $\caloc = \cawall_M$ and $\thetad = \theta_M$. Since the coordinates $(\theta_M , \cawall_M)$ are decided by the limit of $t \to \infty$ from the impulsive time-scale, we directly use the numerical values of $\theta_M$ obtained from our simulations. Using the $C_0$ obtained from numerics for the highest resolution case of figure \ref{fig:ALL_GNBC_SLIP_NO_SLIP_angle_relaxation}, the derived analytical solution \eqref{eqc:sol_ode} shows a good agreement with DNS results in figure~\ref{fig:analytical_vs_numerics}.}
\begin{figure}
    \centering
    \includegraphics[width=0.5\textwidth]{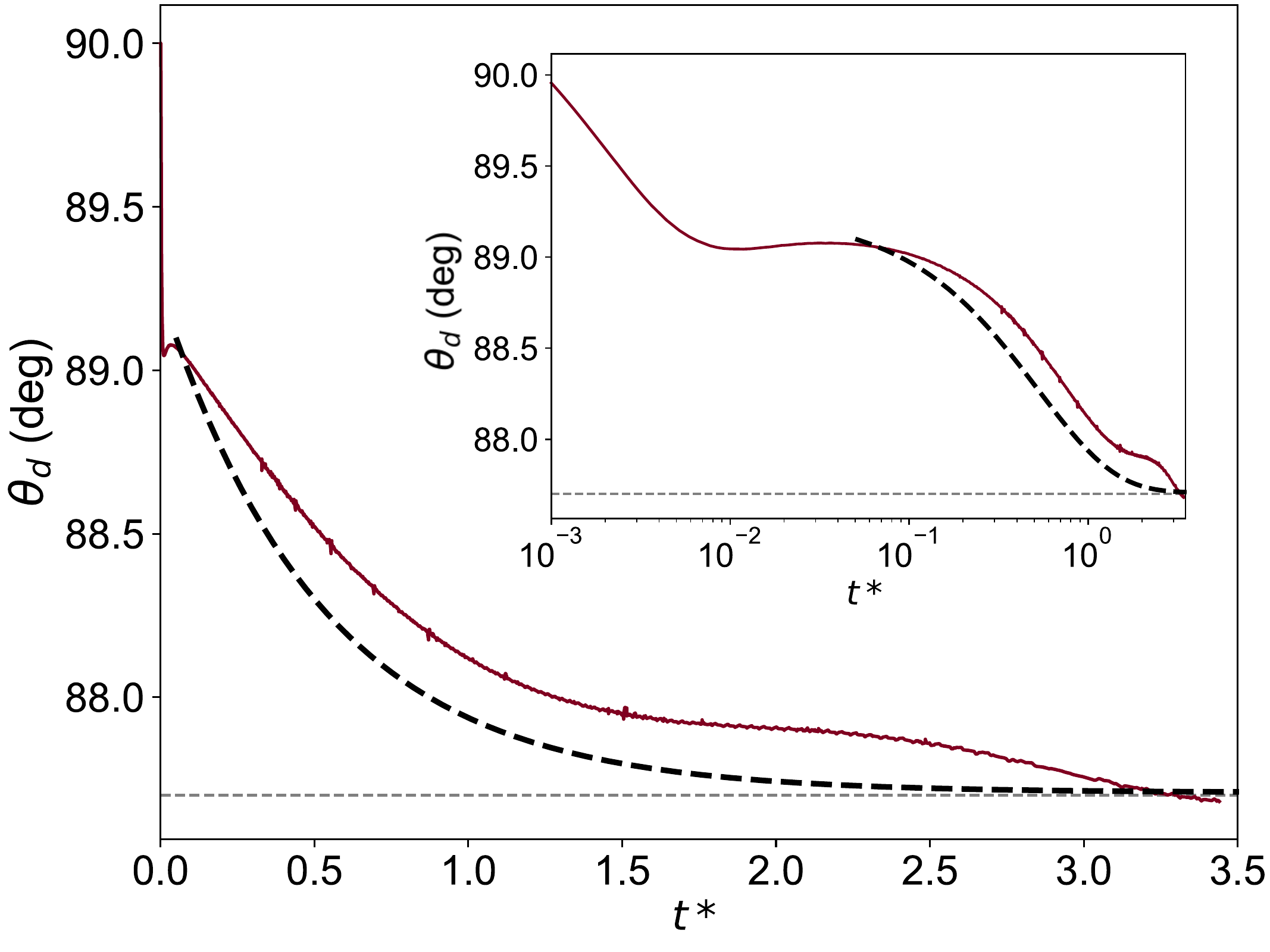}
    \caption{\revgen{Comparison of the dynamic angle $\thetad$ from DNS (red curve) with the analytical solution~\eqref{eqc:sol_ode} (black dashed curve) for the CR-GNBC case with $\cawall = 0.04$, $\varepsilon = 0.05$ and $\varepsilon/\Delta = 40$. Inset shows the same plot in log-scale.}}
    \label{fig:analytical_vs_numerics}
\end{figure}
\FloatBarrier
\backsection[Author contributions]{TF, YK and MF contributed equally to this work that includes free angle method, performing simulations and writing paper with feedback from all authors. Study was performed at all three institutes hosted by SA, DB and SZ. Detailed discussions were done among all authors on all the ideas presented in the paper.}

\backsection[Funding]{This project has received funding from the European Research Council (ERC) under the European Union's Horizon 2020 research and innovation programme (grant agreement n° 883849). MF and DB acknowledge the financial support by the German Research Foundation (DFG) within the Collaborative Research Centre 1194 (Project-ID 265191195). } 

\backsection[Acknowledgments]{MF and DB gratefully acknowledge insightful discussions with Joël De Coninck (ULB).}

\backsection[Declaration of interest]{The authors report no conflict of interest.} 

\appendix
\section{Convergence study}
\label{appex:convergence_study}
\begin{figure}
\centering
     \includegraphics[width=\textwidth]{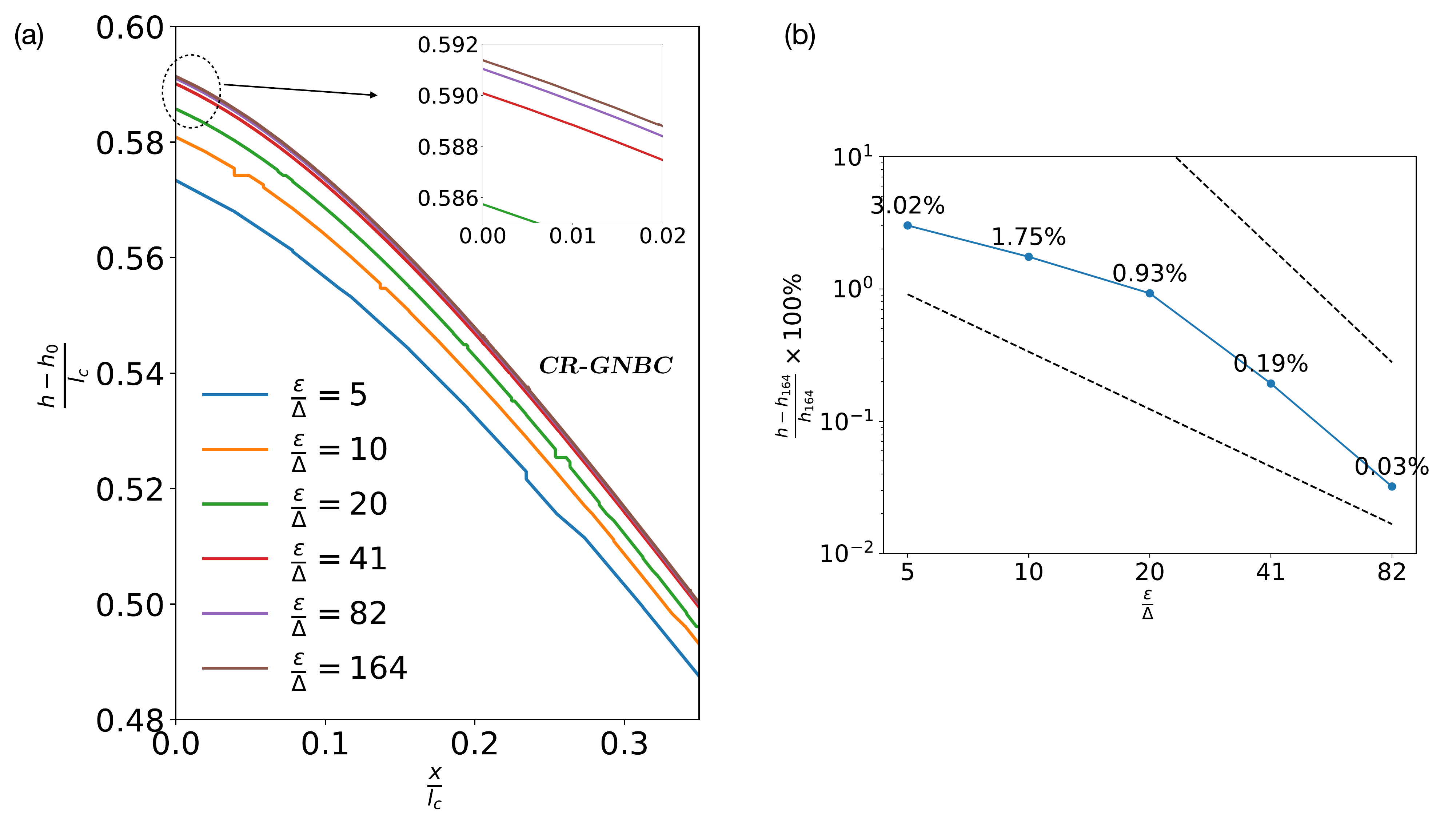}
    \caption{Steady-state height and interface shapes near the contact line with varying grid resolution for the CR-GNBC. In (a), $\cawall=0.12$ and $\varepsilon=0.2$ are fixed, showing convergent interface shapes. In (b), a fixed $\cawall=0.04$ reveals that due to implicit slip, steady-state solutions are achievable even with a no-slip boundary condition. Interface shapes do not converge with grid refinement, and no steady-state height is found at resolutions higher than $\caplength / \Delta > 100$. (b) Percentage error in the contact line position for steady-state interface shapes obtained in (a). The reference solution is taken at 164 grid points per slip length $\varepsilon / \Delta$, and the dashed lines represent second-order and first-order convergence. It is observed that above $20$ grid points per $\varepsilon$, a second-order convergence is achieved.}
\label{fig:ERR_grid_independency_GNBC_No_slip}
\end{figure}


We conduct a convergence study to demonstrate the grid independence of the CR-GNBC. In figure~\ref{fig:ERR_grid_independency_GNBC_No_slip}(a), we present interface shapes for a fixed $\cawall = 0.12$ and $\varepsilon = 0.2$ with varying resolutions, showing apparent convergence. In figure~\ref{fig:ERR_grid_independency_GNBC_No_slip}, we display the percentage error in the contact line position for this case, revealing second-order convergence. This confirms that unlike \cite{Afkhami2018} our CR-GNBC method achieves grid independence for steady-state height with a fixed $\cawall$ and $\varepsilon$, including the $\catr$.

\section{Transition capillary number and the contact region width $\varepsilon$}
\label{appex:ca_tr_eps_GNBC}

\begin{figure}
\centering
    \includegraphics[width=\textwidth]{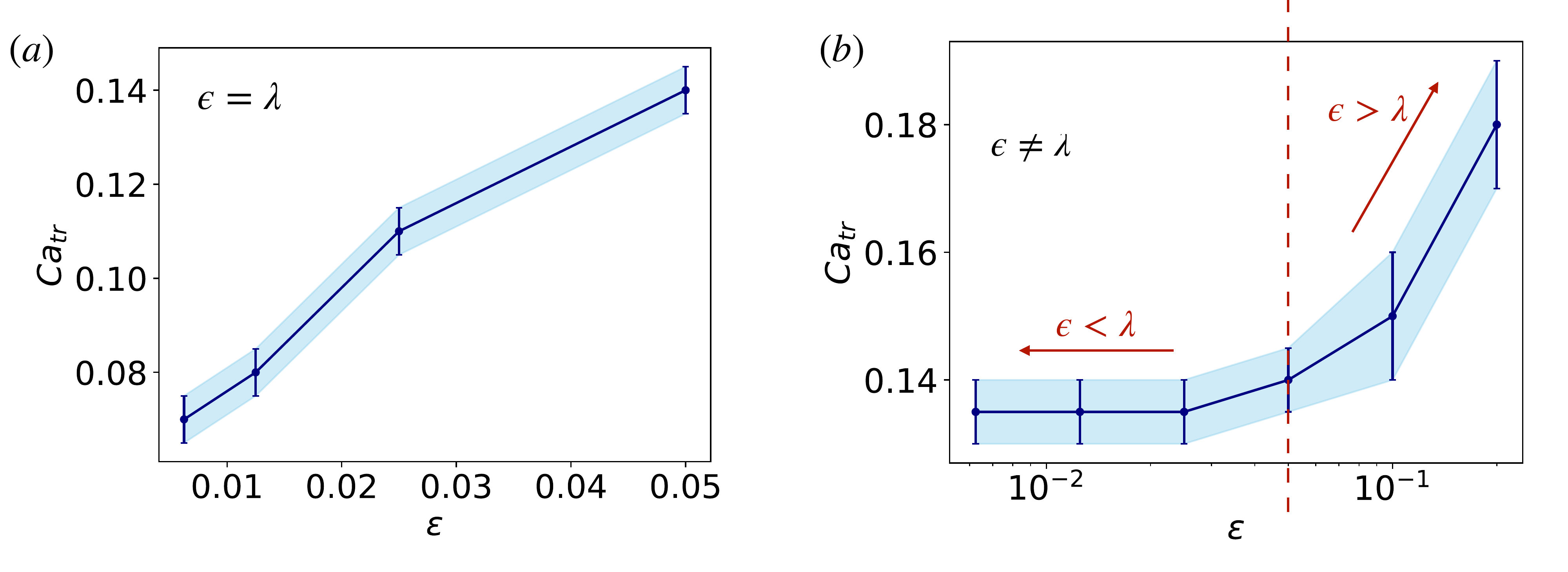}
    \caption{Transition capillary number plotted against variation of (a) $\varepsilon$ and $\lambda$ with $\varepsilon=\lambda$, and (c) $\varepsilon$ such that the slip length $\lambda$ is fixed. All simulations are carried out with the CR-GNBC and $\thetaeq = 90^\degree$. The resolution for all simulations is maintained at $\min(\varepsilon,\lambda) / \Delta = 5.12$.}
    \label{fig:Ca_tr_eps_angle}
\end{figure}

Figure~\ref{fig:Ca_tr_eps_angle} illustrates $\catr$ as a function of $\varepsilon$, $\lambda$ and $\thetaeq$. When we have $\varepsilon=\lambda$, we see that $\catr$ decreases with the decrease of $\varepsilon$. When we break the restriction of $\varepsilon=\lambda$, we see an interesting behaviour in figure~\ref{fig:Ca_tr_eps_angle}\revgen{(b)}. Since we know that $\varepsilon$ and $\lambda$, both promote smoothing behaviour, the $\catr$ is decided by the larger of the two. Hence, when we decrease $\varepsilon$ below the slip length $\lambda$, we see that $\catr$ goes to a constant value. The dependence on the equilibrium contact angle shown in figure~\ref{fig:Ca_tr_eps_angle}(b) is notably linear. While this linearity may break at smaller angles, it's important to note that our solver, which employs only horizontal heights, faces limitations in handling angles smaller than $30^\degree$.

\FloatBarrier

\bibliographystyle{jfm}
\bibliography{bibliography.bib}



\end{document}

%% file: introduction.tex
The phenomenon of dynamic wetting / dewetting requires a relative motion of the contact line, i.e. the triple line at which the liquid-fluid interface and the solid support’s surface meet, against the solid wall. This fundamental process can be modeled in various ways. If the fluid interface and the contact line are modeled as a material surface and a material line, respectively, it is clear that the classical no-slip condition is incompatible with the dynamic wetting process. Mathematically, it has been shown that, for a material interface and contact line, the no-slip boundary condition leads to a discontinuity in the velocity as the contact line is approached. Because of that, a viscous fluid develops a non-integrable singularity at the moving contact line. This has been first shown in the seminal paper by \cite{Huh1971}. Since then, various mathematical models have been developed to resolve the paradox in the continuum mechanical description. In the framework of diffuse interface models, where both the fluid interface and contact line have a finite width characterized by a smooth but rapidly varying order parameter, a motion of the contact line can be achieved by pure diffusion of the order parameter; see \cite{Jacqmin2000}. In this case, the motion is driven by gradients of the chemical potential and the contact line is not a material line with respect to the fluid particles. The Interface Formation Model due to \cite{Shikhmurzaev1993,Shikhmurzaev2008} describes the dynamic wetting process using mass transfer between the bulk phases of the liquid and the interfaces between fluid and solid and fluid and gas. Hence, in this case, the contact line can move without hydrodynamic slip, \revgen{displaying a rolling motion.} \\
\\
A commonly used approach in the sharp interface framework is to model the interface and the contact line as material objects and to allow for slip between the bulk fluid and the solid wall. The Navier slip condition states that the amount of tangential slip is determined by a balance between the tangential component of the viscous stress, described by the viscous stress tensor $\mat{S}$, and a sliding friction force between fluid particles and the solid surface according to 
\begin{align}\label{eqn:intro-navier-slip}
- \beta (\mathbf{v}_\parallel-\vuwall) = (\mat{S} \ndomega)_\parallel,
\end{align}
\revthree{where $\ndomega$ denotes the unit outer normal to the solid boundary.} This boundary condition introduces the slip length \revgen{$\lambda:=\eta/\beta$} as the key parameter, where $\eta$ denotes the viscosity of the liquid and $\beta > 0$ is a coefficient describing the (sliding) friction between the liquid molecules and the solid surface. Within the Navier slip model, the slip length can be interpreted geometrically as the distance below the solid surface where the linearly extrapolated tangential velocity vanishes. It is well-known that the singularity at the moving contact line is only partially relaxed by the Navier slip condition. A logarithmic divergence as a function of the distance to the contact line still exists for the curvature and the pressure, as pointed out by \cite{Huh1977}. However, the singularity is transformed into an integrable one and, hence, physically meaningful solutions are possible, at least for the macroscopic flow. The physical implications of the pressure singularity are debated in the literature. \cite{Shikhmurzaev2006} argues that the pressure should remain finite because otherwise the model of an incompressible fluid would no longer be valid. On the other hand, it has been demonstrated that the Navier slip model is able to describe various wetting experiments in a satisfactory manner, \revgen{see ~\cite{Fullana2020}.}\\
Besides the mobility of the contact line, the wettability of the solid surface is another key parameter for the physical system. It is usually characterized by the equilibrium contact angle $\thetaeq$ that the fluid interface forms with the solid boundary in equilibrium. It can be computed from the surface tension of the liquid-gas, liquid-solid and solid-gas interfaces, using the equation introduced by \cite{Young.1805}, viz.
\begin{align}
\sigma \cos \thetaeq + {\sigma_{\text{ls}}} - {\sigma_{\text{sg}}} = 0.  
\end{align}
While the latter equation can be easily deduced from variational principles, the dynamics of the contact angle is a much more complex problem and a large variety of empirical models exist. Notably, there is one fundamental relation for the dynamics of the microscopic contact angle $\thetad$ in the limit of slow velocities of the contact line, which is shared by many of these models:
\begin{align}\label{eqn:linear_response_theory}
- \zeta \clspeed = \sigma (\cos \thetad - \cos \thetaeq). 
\end{align}
Here $\clspeed$ denotes the normal speed of the contact line relative to the solid surface (positive for advancing and negative for a receding contact line) and $\zeta$ is the so-called ``contact line friction'' parameter. Equation \eqref{eqn:linear_response_theory} arises, for example, from the molecular kinetic theory of wetting in the limit of small capillary numbers, i.e.\ for a slow motion of the contact line (see \cite{Blake1969,Blake2015}).\\
\\
Recently, \cite{Fricke2019,Fricke2018} studied the fundamental kinematics of the contact angle transport and showed that the rate-of-change of the contact angle is fully determined by a directional derivative of the velocity field $\mathbf{v}$ at the contact line, viz.\
\begin{align}\label{eqn:contact-angle-evolution-equation}
\dthetad = (\partial_{\mathbf{\tau}} \mathbf{v}) \cdot \nsigma.
\end{align}
Here $\nsigma$ denotes the interface normal vector and $\mathbf{\tau}$ is the unit vector tangential to the interface \revgen{and normal to the contact line} (see Section~\ref{sec:mathematical_model} for more details). \secondrev{Notably, when applied to the full two-phase flow problem (assuming sufficient regularity of the solution) Equation~\eqref{eqn:contact-angle-evolution-equation} can be used to show that $\dthetad$ is proportional to the tangential component of the shear stress that also appears in the Navier slip condition (see \eqref{eqn:theta_dot_relation} in Section \ref{sec:mathematical_model}). In particular, this shear stress component vanishes if the contact angle does not change in time, i.e.\ in a quasi-steady state.} In other words, \eqref{eqn:contact-angle-evolution-equation} predicts an ``apparent perfect slip'' at the moving contact line if $\dthetad=0$. Indeed, indications of a vanishing shear stress in the vicinity of the contact line have been observed in molecular dynamics (MD) simulations by \cite{Thompson1989} and others. We will see that perfect slip in the sense of vanishing shear stress is possible within the GNBC model, which makes the model consistent with equation~\eqref{eqn:contact-angle-evolution-equation}. On the other hand, \cite{Fricke2019} and \cite{Fricke2020a} showed that the Navier slip model \eqref{eqn:intro-navier-slip} with a contact angle boundary condition like \eqref{eqn:linear_response_theory} is not consistent with \eqref{eqn:contact-angle-evolution-equation} and regular solutions, if they exist, show an unphysical behavior.\\
\\
The ``Generalized Navier Boundary Condition'' (GNBC) was first described by \cite{Qian2003,Qian2006,Qian2006a} in the context of diffuse interface models \revgen{and molecular dynamics}. The key idea of the GNBC is to introduce the \emph{uncompensated Young stress} as an additional force density into the constitutive relation \eqref{eqn:intro-navier-slip}. So, in this model, the slip velocity relative to the solid surface is a result of a balance between a sliding friction force, the viscous stress and the uncompensated Young stress. In a sharp interface and sharp contact line formulation, the GNBC can be written as (see \revone{\cite{Ren2007}}, \cite{Gerbeau2009})
\begin{align}
\label{eqn:intro/formal_gnbc}
-\beta (\mathbf{v}_\parallel - \vuwall) = (\mat{S}\ndomega)_\parallel + \sigma (\cos \thetad - \cos \thetaeq) \, \ngamma \delta_\Gamma \quad \text{on} \quad \partial\Omega.
\end{align}
Notably, the contact line delta distribution $\delta_\Gamma$ appears because, in the sharp contact line formulation, the Young stress is concentrated on a mathematical curve. Hence, Equation~\eqref{eqn:intro/formal_gnbc} should mathematically be understood as an equality of distributions. This delta function GNBC formulation is applicable in weak formulations of the two-phase flow problem where the contact line delta distribution will translate into an integral over the contact line in the weak formulation (see, e.g, \cite{Gerbeau2009,Fumagalli2018}). On the other hand, there is no contact line delta distribution in the Phase Field formulation of the GNBC due to Qian et al., because the thickness of the interface and the contact line is a finite, physical model parameter in this case. \cite{Yamamoto2013,Yamamoto2014} implemented the GNBC approach into a front-tracking-method and studied the dynamics of capillary rise in a tube. In this method, the contact line is transported by advecting the Lagrangian marker points without a prescribed contact angle. Then, the dynamic contact angle is evaluated and used to compute the uncompensated Young stress, which determines the slip velocity profile. Yamamoto et al.\ noticed that the viscous stress becomes negligible as the contact line is approached. Motivated by this observation, they dropped the viscous stress contribution in \eqref{eqn:intro/formal_gnbc} leading to a ``simplified GNBC'', formally reading as
\begin{align}
\label{eqn:intro/simplified_gnbc}
-\beta (\mathbf{v}_\parallel - \vuwall) = \sigma (\cos \thetad - \cos \thetaeq) \, \ngamma \delta_\Gamma.
\end{align}
It is evident that, by taking the inner product with the contact line normal vector, Eq.~\eqref{eqn:intro/simplified_gnbc} can be formally reduced to an equation equivalent to \eqref{eqn:linear_response_theory} if the delta distribution is approximated with a regular function over a finite width. Indeed, \cite{Yamamoto2013,Yamamoto2014} smoothed the delta distribution over a region of approximately four grid points. Using this estimate as the characteristic width of the delta function, the authors concluded that
\begin{align}\label{eqn:intro/cl-speed-Yamamoto}
\clspeed \approx \frac{\lambda}{\Delta} \frac{\sigma (\cos \thetaeq - \cos \thetad)}{\eta}, 
\end{align}
where $\Delta$ is the grid size. Obviously, the contact line speed in \eqref{eqn:intro/cl-speed-Yamamoto} can only be grid-independent if also the slip length is chosen in proportion to the grid size, i.e.\ if $\lambda \sim \Delta$. Consequently, they fixed the parameter $\chi:=\lambda/\Delta$ in their simulations. The approach was extended by using the Cox-Voinov relation for $\thetad$ in \cite{Yamamoto2014}. Later, \cite{Yamamoto2016} used this method to study the withdrawing plate problem with a single wettable defect. Recently, the GNBC front-tracking approach was extended by \cite{Kawakami2023} using a so-called ``rolling belt-model'' inspired by the work of \cite{Lukyanov2017}. \cite{Chen2019} used the GNBC in a Front Tracking method to study the coalescence-induced self-propelled motion of droplets on a solid surface. \cite{Shang2018} used a quite similar method to study droplet spreading and the motion of drops on surfaces subject to a shear flow. All these methods have in common that the uncompensated Young stress is distributed over a characteristic distance, which is related to the mesh size, \revgen{see~\cite{Ren2007,Ren2011b,Ren2011,ZHANG2020-JCP}.}\\
\\
In the present work, we propose a ``sharp-interface, contact region GNBC'' (CR-GNBC) formulation, where the contact line delta distribution is replaced by a smooth function with a characteristic width $\varepsilon > 0$. This width $\varepsilon$ is understood as a physical model parameter and is, therefore, chosen independently of the computational mesh. It has been shown recently by \cite{Kulkarni2023} that this model (i.e.\ the GNBC model with finite $\varepsilon$) admits a local $\mathcal{C}^2$-regularity of the velocity in the vicinity of the moving contact line. We develop an implementation of the CR-GNBC in a geometrical Volume-of-Fluid method. This method turns out to be consistent with the fundamental kinematic law \eqref{eqn:contact-angle-evolution-equation}. For this purpose, the dynamic contact angle is not prescribed but is reconstructed from the volume fraction field in a neighborhood of the contact line. As one important preliminary step, we validate this ``free contact angle'' method by studying the advection problem by a \emph{prescribed} velocity field (see \cite{Fricke2020}). In this case, the interface and the contact line are transported without a boundary condition for the contact angle and the results are validated against analytical solutions of \eqref{eqn:contact-angle-evolution-equation}. We couple this method to the CR-GNBC model and use the reconstructed contact angle $\thetad$ as an input parameter to compute the uncompensated Young stress in the simulation.

\subsection*{Structure of this article}
\begin{figure}
\centering
\includegraphics[width=0.5\columnwidth]{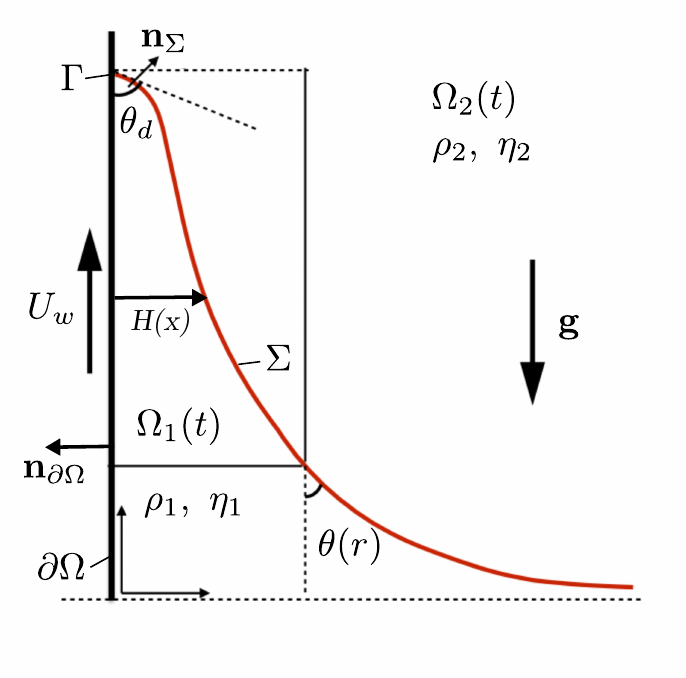} 
\caption{Mathematical notation for the withdrawing tape setup.}\label{fig:notation}
\end{figure}
We study the withdrawing tape problem as a prototypical example for a dynamic dewetting process. The setup follows the previous work by \cite{Afkhami2018}. The solid wall is moving upwards with velocity $\uwall \geq 0$ in the laboratory frame (see figure~\ref{fig:notation}). So, we study the case of a receding contact line. We define the (global) capillary number with respect to the wall speed as
\begin{align}
\label{eqn:intro/cawall}
\cawall := \frac{\eta \uwall}{\sigma}.
\end{align}
For convenience, we define the \emph{contact line} capillary number using the negative contact line speed, i.e.\ (note that the contact line speed $\clspeed$ is always measured relative to the solid)
\begin{align}
\label{eqn:intro/cacl}
\cacl := \frac{\eta (-\clspeed)}{\sigma}.
\end{align}
In a quasi-stationary state, we have $-\clspeed = \uwall$ and hence $\cawall = \cacl$. With this definition, we can always work with positive values for the capillary number. Notice that, in the literature, one will also find the convention that $\cacl$ is negative for a receding contact line and positive for an advancing contact line.\\
\\
The mathematical derivation of the GNBC in a sharp-interface framework is revisited in Section~\ref{sec:mathematical_model}. It is shown that the GNBC can be obtained as a combined closure relation for the dissipation due to slip along the liquid-solid surface and the contact line dissipation. Using the laws of kinematics, we derive the contact angle evolution equation in Section~\ref{subsection:contactAngle_evolution} and show that \eqref{eqn:linear_response_theory} holds for quasi-stationary states. Moreover, the GNBC thin film equation is derived in Section~\ref{subsection:thin_film}. The numerical implementation of the method in the geometrical Volume-of-Fluid solver is described in Section~\ref{sec:Numeric_method}. We validate the numerical method by studying the kinematic transport of the contact angle and the curvature at the contact line. The results for the withdrawing tape problem are discussed in detail in Section~\ref{sec:Results}. In particular, it is shown that the results are converging under mesh refinement. Notably, we can demonstrate by a mesh study that, unlike for the Navier slip model, the curvature at the contact line converges to a finite value. Finally, we conclude this study by an outlook to a non-linear variant of the GNBC, which can be derived as a non-linear closure based on the entropy production described earlier in Section~\ref{sec:mathematical_model}.

\begin{table}
\centering
\caption{List of Symbols}
\begin{tabular}{ccc}
\toprule
Symbol & Description & Units \\
\midrule
$\rho$ & Density & kg/m\textsuperscript{3} \\
$\visc$ & Viscosity & Pa·s \\
$\sigma$ & Surface Tension & N/m \\
$\mathbf{v}$ & Velocity & m/s \\
$\beta$ & Friction Coefficient & Pa·s/m \\
$\mathbf{g}$ & Gravitational Acceleration & m/s\textsuperscript{2} \\
$\mat{D}$ & Rate-of-Deformation Tensor & 1/s \\
$\mat{S}=2\visc \mat{D}$ & Viscous Stress Tensor & Pa \\
\addlinespace[1em]
$\Sigma$ & Interface & - \\
$\normalspeed$ & Interface Normal Speed & m/s \\
$\nsigma$ & Unit Interface Normal & - \\
$\kappa = - \nabla_\Sigma \cdot \nsigma$ & Interface Mean Curvature & 1/m \\
$\Gamma$ & Contact Line & - \\
$\clspeed$ & Contact Line Speed & m/s \\
$\ngamma$ & Contact Line Normal (tangential to $\partial\Omega$) & - \\
$\partial\Omega$ & Solid Boundary & - \\
$\ndomega$ & Unit Outer Normal to $\Omega$ & - \\
$\uwall$ & Wall Speed & m/s \\
$\vuwall$ & Wall Velocity & m/s \\
\addlinespace[1em]
$\lambda=\visc/\beta$ & Slip Length & m \\
$\cawall$ & Wall Capillary Number & - \\
$\cacl$ & Contact Line Capillary Number & - \\
$\caloc$ & Capillary Number in the lab frame of reference & - \\
$\catr$ & Transition Capillary Number & - \\
$\zeta$ & Contact Line Friction & Pa·s \\
$\thetaeq$ & Static Contact Angle & rad \\
$\thetad$ & Dynamic Microscopic Contact Angle & rad \\
\revgen{$\thetaabove$} & \revgen{Numerical angle one layer above the wall} & rad \\
$\thetasteady$ & Steady state contact angle & rad \\
\\
\revgen{$t^*$} & \revgen{Dimensionless time} & \revgen{-} \\
\revgen{$t_{\varepsilon}$} & \revgen{Slip length time scale} & \revgen{-} \\
\revgen{$t_{r}$} & \revgen{Relaxation time scale} & \revgen{-} \\
\bottomrule
\end{tabular}
\end{table}

%% file: modeling.tex
\subsection{Governing equations}
We employ the sharp-interface continuum modeling approach. We start from the incompressible, two-phase Navier-Stokes equations with surface tension for Newtonian fluids under isothermal conditions (see, e.g., \cite{Slattery.1999,Pruess2016}).  Inside the fluid phases, the governing equations are
\begin{align}
\partial_t (\rho \mathbf{v}) + \nabla \cdot (\rho \mathbf{v} \otimes \mathbf{v}) + \nabla p &= \nabla \cdot \mat{S} + \rho \mathbf{g}, \label{eqn:bulk-eq-1} \\
\quad \nabla \cdot \mathbf{v} &= 0 \label{eqn:bulk-eq-2}
\end{align}
with the viscous stress tensor\footnote{We use the symbol $\mat{D}=\frac{1}{2} (\nabla \mathbf{v} + \nabla \mathbf{v}^\transpose)$ for the rate-of-deformation tensor.}
\[ \mat{S} = 2 \eta \mat{D} = \eta (\nabla \mathbf{v} + \nabla \mathbf{v}^\transpose). \]
These bulk equations are accompanied by jump conditions at the interface $\Sigma(t)$. The interface is modeled as a hypersurface (i.e.\ it has zero thickness) and separates the domain $\Omega$ into two bulk phases $\Omega_{1,2}(t)$ occupied by the two fluid phases (see figure~\ref{fig:notation}). Assuming that no phase change occurs in the system, the normal component of the adjacent fluid velocities $\mathbf{v}_{1,2}$ at the interface are coinciding and equal to the speed of normal displacement $V_\Sigma$ of the interface, resulting in the kinematic boundary condition
\begin{align}
 V_\Sigma = \mathbf{v} \cdot \nsigma \quad &\text{on} \quad \Sigma(t),
\end{align}
where $\nsigma$ is the interface unit normal field. Additionally, no-slip between the fluid phases is usually assumed. Assuming further that the surface tension $\sigma$ is constant, the jump conditions for mass and momentum read as
\begin{align}\label{eqn:interface_jump_conditions}
\jump{\mathbf{v}} = 0, \quad \jump{p \mat{\mathds{I}} - \mat{S}} \nsigma = \sigma \kappa \nsigma \quad \text{on} \quad \ \Sigma(t).
\end{align}
Here $\kappa := - \nabla_\Sigma \cdot \nsigma$ is twice the mean curvature of the interface and 
\[ \jump{\psi}(t,\mathbf{x}) := \lim_{h \rightarrow 0^+} (\psi(t,\mathbf{x}+h\nsigma) - \psi(t,\mathbf{x}-h\nsigma)) \]
is the jump of a quantity $\psi$ across the interface. We assume that the solid boundary $\partial\Omega$ is not able to store mass and we assume it to be impermeable. We consider an inertial frame of reference, where the wall is moving parallel to itself with a velocity $\uwall \geq 0$ upwards (see figure~\ref{fig:notation}). The impermeability condition in this frame of reference reads as
\begin{align}\label{eqn:impermeability_condition}
\mathbf{v}_\bot = 0 \quad &\text{on} \ \partial\Omega,
\end{align}
where $\mathbf{v}_\bot = (\mathbf{v} \cdot \ndomega) \, \ndomega$ denotes the normal part of the velocity with respect to $\partial\Omega$.\newline
\newline
In order to obtain a closed model, the system of equations \eqref{eqn:bulk-eq-1}-\eqref{eqn:impermeability_condition} must be complemented by (one or more) additional boundary conditions describing
 \begin{enumerate}[(1)]
  \item the \textbf{wettability} of the solid (i.e.\ the static and dynamic contact angle) and
  \item the \textbf{mobility} of the contact line (i.e.\ the tangential velocity $\mathbf{v}_\parallel$ at the solid boundary).
  \end{enumerate}
These boundary conditions are closure relations for the continuum mechanical description and must be thermodynamically consistent, i.e.\ they must obey in particular the second law of thermodynamics. To arrive at a consistent closure, we consider the available energy consisting of the kinetic energy of the bulk phases and the surface energies of the liquid-gas interface as well as the wetted area $W(t) \subset \partial\Omega$, i.e.\
\[ E(t) := \int_{\Omega\setminus\Sigma(t)} \frac{\rho |\mathbf{v}|^2}{2} \, dV + \int_{\Sigma(t)} \sigma \, dA + \int_{W(t)} \sigma_w \, dA. \]
Here $\sigma = \sigma_{\text{lg}} > 0$ denotes the surface tension of the liquid-gas interface and
\[ \sigmawet = \sigma_{\text{ls}} - \sigma_{\text{sg}}  \]
is the specific energy density for wetting the solid surface. Note that $\sigmawet$ \revthree{is negative for a hydrophilic surface}, as we see from Young's equation 
\begin{align}
\sigma \cos \thetaeq + \sigmawet = 0,  
\end{align}
which defines the ``static'' or ``equilibrium'' contact angle $\thetaeq$. It is a purely mathematical exercise \revone{(see \cite{Ren2007,Ren2011}, \cite{Fricke2021} for details)} to compute the rate of change $\dot{E}$ for a sufficiently regular solution of \eqref{eqn:bulk-eq-1}-\eqref{eqn:impermeability_condition} (in the absence of external forces, i.e.\ for $\mathbf{g}=0$). The result reads as
\begin{align}
\label{eqn:energy_dissipation_semiclosed_model}
\ddt{E} = - \int_{\Omega\setminus\Sigma(t)} \mat{S}:\mat{D} \, dV + \int_{\partial\Omega} (\mathbf{v}_\parallel-\vuwall) \cdot (\mat{S}\ndomega)_\parallel \, dA +  \int_{\Gamma(t)} \sigma(\cos \thetad - \cos \thetaeq) \, \clspeed \, dl.
\end{align}
In this formulation with a continuous velocity field, the scalar contact line speed (measured relative to the solid) is given as
\[  \clspeed = \mathbf{v} \cdot \ngamma-\uwall. \]
Closure relations are required to satisfy the second law of thermodynamics\footnote{Note that we assume an isothermal system here. In this case, we may directly consider the change in available energy.} $\dot{E} \leq 0$. The first contribution in \eqref{eqn:energy_dissipation_semiclosed_model} has a negative sign as we consider incompressible Newtonian fluids, i.e. $\mat{S}=2\eta \mat{D}$. A linear closure for the second integral in \eqref{eqn:energy_dissipation_semiclosed_model} yields the well-known Navier slip condition, i.e.
\begin{align}\label{eqn:navier_slip_condition}
-\beta (\mathbf{v}_\parallel-\vuwall) = (\mat{S}\ndomega)_\parallel \quad \text{with a friction coefficient} \quad \beta \geq 0.
\end{align}
Using the slip length parameter $\lambda=\eta/\beta$, one may reformulate \eqref{eqn:navier_slip_condition} as
\begin{align}\label{eqn:navier_slip_condition_v2}
\mathbf{v}_\parallel + 2 \lambda (\mat{D}\ndomega)_\parallel = \vuwall.
\end{align}
The third integral in \eqref{eqn:energy_dissipation_semiclosed_model} suggests that the dynamic contact angle $\thetad$, which is \emph{mathematically} defined as the angle of intersection\footnote{Note that, in order to define the contact angle $\thetad$, we have to assume that interface $\Sigma(t)$ has a well-defined normal field up to the boundary. This is the case even if the curvature has a logarithmic, hence integrable, singularity.} of the free surface $\Sigma$ with the solid boundary $\partial\Omega$, i.e.\
 \[ \cos \theta_d := - \nsigma \cdot \ndomega \quad \text{at} \ \Gamma(t), \]
should be linked to the contact line speed $\clspeed$. A linear closure leads to the well-known condition
\begin{align}\label{eqn:linear_contact_angle_model}
 -\zeta \clspeed = \sigma (\cos \thetad - \cos \thetaeq) \quad \text{with} \quad \zeta \geq 0,
\end{align}
where $\zeta$ can be seen as a contact line friction coefficient. Note that also more general contact angle boundary conditions are possible if a non-linear closure relation is employed. To summarize, the ``standard model''\footnote{The mathematical model \eqref{eqn:bulk-eq-1}-\eqref{eqn:impermeability_condition}, \eqref{eqn:navier_slip_condition_v2}, \eqref{eqn:linear_contact_angle_model} is one of the most commonly applied models for dynamic wetting in the literature. However, there are many more modeling approaches which aim at a regularization of the singularity and a prediction of the dynamics of wetting. For a survey of the field, we refer to the references \cite{Gennes2004,Blake2006,Shikhmurzaev2008,Bonn2009,Snoeijer2013,Marengo2022}.} based on the Navier slip condition is given by Equations~\eqref{eqn:bulk-eq-1}-\eqref{eqn:impermeability_condition}, \eqref{eqn:navier_slip_condition_v2} together with \eqref{eqn:linear_contact_angle_model} or a non-linear variant of the form
  \begin{equation}\label{eqn:nonlinear-contact-angle-bc}
 \begin{aligned}
\thetad = f(\clspeed) \quad &\text{on} \quad \Gamma(t).
 \end{aligned}
 \end{equation}
 To ensure thermodynamic consistency, we require that
 \[ \eta \geq 0, \ \ \sigma \geq 0, \ \ \lambda \geq 0, \ \ \clspeed (f(\clspeed)-\thetaeq) \geq 0. \]
As shown by \cite{Fricke2020a}, \revgen{for regular solutions,} there is \revgen{a fundamental inconsistency of boundary conditions in the standard model. In fact,} the evolution of the contact angle is determined by the contact angle boundary condition (say \eqref{eqn:linear_contact_angle_model}) \emph{as well as} by the flow in the vicinity of the contact line according to \eqref{eqn:contact-angle-evolution-equation}. As a consequence, a regular solution of the system does not exist but a weak singularity is present at the contact line as shown already in \cite{Huh1977}. 

\subsection{Formal derivation of the GNBC}\label{sec:gnbc_derivation} 
\revgen{We now recall the derivation of the GNBC in the context of the sharp-interface framework described above starting from \eqref{eqn:energy_dissipation_semiclosed_model}}\\

It is important to note that the  GNBC was originally formulated in a \emph{diffuse} interface framework (see \cite{Qian2003,Qian2006}).  However, the GNBC can be formally understood in the sharp interface model as a \emph{combined closure} for the terms in the entropy production \eqref{eqn:energy_dissipation_semiclosed_model} that arise from the contact line motion and from slip at the solid-liquid boundary. The combined closure \revgen{formally leads} to a single boundary condition instead of the two independent conditions in the standard Navier slip model. Hence, the number of boundary conditions is reduced and one can show that the inconsistency at the contact line \revgen{can be resolved this way} (see below).\\
\\
As a starting point, we consider the sum of the wall and the contact line dissipation, given as
\begin{align*} 
\mathcal{T} = \int_{\partial\Omega} (\mathbf{v}_\parallel-\vuwall)\cdot(\mat{S}\ndomega)_\parallel \, dA +  \sigma  \int_{\Gamma(t)} (\cos \thetad - \cos \thetaeq) \, \clspeed \, dl. 
\end{align*}
By introducing the contact line delta distribution $\delta_\Gamma$, one can rewrite $\mathcal{T}$ as a single integral over the entire solid boundary $\partial\Omega$ according to
\begin{equation*}
\begin{aligned}
\mathcal{T} = \int_{\partial\Omega} \left( (\mat{S}\ndomega)_\parallel + \sigma (\cos \thetad - \cos \thetaeq) \, \ngamma \delta_\Gamma \right) \cdot (\mathbf{v}_\parallel-\vuwall) \, dA.
\end{aligned}
\end{equation*}
Note that it is possible to factor out the common co-factor $(\mathbf{v}_\parallel-\vuwall)$ because the contact line speed can be written as $\clspeed = (\mathbf{v}_\parallel-\vuwall) \cdot \ngamma$. A \emph{linear}\footnote{A non-linear generalization of the closure is discussed in Section~\ref{sec:non-linear-gnbc}.} closure relation is now provided by the generalized Navier boundary condition
\begin{align}
\label{eqn;mathematical_models/formal_gnbc}
-\beta (\mathbf{v}_\parallel - \vuwall) = (\mat{S}\ndomega)_\parallel + \sigma (\cos \thetad - \cos \thetaeq) \, \ngamma \delta_\Gamma \quad \text{on} \quad \partial\Omega
\end{align}
with a friction coefficient $\beta > 0$. Notice that the boundary condition given by \eqref{eqn;mathematical_models/formal_gnbc}, called  ``delta function GNBC'' to distinguish it from other variants, should  be understood in the sense of distributions.\\
\\
\revone{There is an interesting link between \eqref{eqn;mathematical_models/formal_gnbc} and the independent closure \eqref{eqn:navier_slip_condition_v2} together with \eqref{eqn:linear_contact_angle_model}: following \cite{Ren2007}, one may assume that the friction $\beta$ in \eqref{eqn;mathematical_models/formal_gnbc} has a singular component, i.e.\ 
\[ \beta = \beta_\Gamma \delta_\Gamma + \beta_b \quad \text{with} \quad \beta_\Gamma, \beta_b \geq 0. \]
In this case, equation \eqref{eqn:navier_slip_condition_v2} takes the form
\[ [\beta_\Gamma (\mathbf{v}_\parallel - \vuwall) + \sigma (\cos \thetad - \cos \thetaeq) \, \ngamma] \delta_\Gamma  + [\beta_b (\mathbf{v}_\parallel - \vuwall) + (\mat{S}\ndomega)_\parallel] = 0 \quad \text{on} \quad \partial\Omega. \]
This distributional equation can be split into a singular and a regular part.  This yields the "standard closure relations" for slip and dynamic contact angle, i.e.
\[ \quad \beta_b (\mathbf{v}_\parallel - \vuwall) + (\mat{S}\ndomega)_\parallel = 0, \quad \beta_\Gamma (\mathbf{v}_\parallel - \vuwall) \cdot \ngamma + \sigma (\cos \thetad - \cos \thetaeq) = 0.  \]
Hence, the sharp-interface, sharp-contact line GNBC condition \eqref{eqn;mathematical_models/formal_gnbc}  and the standard closure \eqref{eqn:navier_slip_condition_v2} and \eqref{eqn:linear_contact_angle_model} are formally equivalent if the friction $\beta$ has a singular component at the contact line. As we will see below, this singular component $\beta_\Gamma$ may arise as a "lumped friction" that results from shrinking the contact zone to a line with vanishing thickness. Let us already note at this point, that we are not going to study the limit of a sharp contact line but model it as a finite contact region.} 

\subsection{Contact Region GNBC (CR-GNBC) model}
\label{subsec:Delta_func_approx_GNBC_mobility}
To obtain the CR-GNBC model, we replace the contact line delta distribution in \eqref{eqn;mathematical_models/formal_gnbc} by a smooth function defined over a finite transition region with characteristic width $\varepsilon$ such that
\[ \delta_\Gamma^\varepsilon \geq 0, \quad \int \delta_\Gamma^\varepsilon(x) \, \text{d}x = 1. \]
Note that this approach also requires to extend the definition of the contact angle $\thetad$ and the contact line normal $\ngamma$ away from the sharp contact line. In fact, the existence of the solid boundary, touching the interface at an angle strictly between $0$ and $\pi$, provides a means for this extension of the contact line to a finite region. Then, the deviation of the contact angle from the equilibrium value appears in the velocity boundary condition leading to a force balance between sliding friction forces due to slip along the solid boundary, the tangential component of the viscous stress at the boundary and the uncompensated Young force:
\begin{align}
\label{eqn;mathematical_models/regularized_gnbc}
- (\mathbf{v}_\parallel-\vuwall) = 2 \lambda (\mat{D}\ndomega)_\parallel + \frac{\sigma}{\beta} [(\cos \thetad - \cos \thetaeq) \, \ngamma  \delta^\varepsilon_\Gamma].
\end{align}
Notably, the dynamic contact angle is not prescribed explicitly in this approach. Instead, the dynamics of the contact angle is determined by \eqref{eqn;mathematical_models/regularized_gnbc} and the kinematics of the interface transport.


\subsection{Kinematics of the dynamic contact angle}\label{subsection:contactAngle_evolution}
We derive the evolution law for the contact angle, given a sufficiently regular solution of the CR-GNBC model \eqref{eqn:bulk-eq-1}-\eqref{eqn:impermeability_condition} and \eqref{eqn;mathematical_models/regularized_gnbc}. Below, we consider the limiting case of a free surface flow, where one phase is assumed to be a dynamically passive gas at a constant pressure. We assume that $\delta_\Gamma^\varepsilon$ evaluated at the contact line yields a value of $1/\varepsilon$. Then, the CR-GNBC condition, evaluated at the contact line, reads as
\begin{align}
\label{eqn:kinematics/regularized_gnbc}
\beta (\mathbf{v}_\parallel-\vuwall) + (\mat{S}\ndomega)_\parallel + \frac{1}{\varepsilon} \sigma (\cos \thetad - \cos \thetaeq) \, \ngamma = 0 \quad \text{at} \ \Gamma.
\end{align}
By taking the inner product with the contact line normal vector $\ngamma$ (normal to the contact line and tangential to the solid), we obtain the relation
\begin{align}
\label{eqn:kinematics/regularized_gnbc_v2}
\frac{\clspeed}{\lambda} + \inproduct{\ngamma}{(\nabla \mathbf{v}) \, \ndomega} + \inproduct{(\nabla \mathbf{v}) \, \ngamma}{\ndomega} + \frac{\sigma}{\varepsilon \visc} (\cos \thetad - \cos \thetaeq) = 0 \quad \text{at} \ \Gamma.
\end{align}
Using the kinematic evolution equation for the contact angle derived in \cite{Fricke2019}, one can show that the rate-of-change of the contact angle $\dthetad$ is given as
\begin{align}\label{eqn:theta_dot_relation}
2 \dthetad = - \inproduct{\ngamma}{(\nabla \mathbf{v}) \, \ndomega}.
\end{align}
Moreover, it follows from the impermeability condition that the term $\inproduct{\nabla \mathbf{v} \, \ngamma}{\ndomega}$ vanishes for a flat solid boundary. Therefore, we obtain the contact angle evolution law for a regular solution of the CR-GNBC model. It reads as
\begin{align}
\label{eqn:kinematics/theta_evolution_gnbc}
\dthetad = \frac{\clspeed}{2 \lambda} + \frac{1}{\varepsilon} \, \frac{\sigma}{2\eta} (\cos \thetad - \cos \thetaeq).
\end{align}

\subsection*{Remarks}
\begin{enumerate}[(i)]
 \item Compared to the standard Navier slip model (see \cite{Fricke2019} for details), the uncompensated Young stress leads to an additional term in the equation for $\dthetad$, which reads as
\[ \frac{1}{\varepsilon} \, \frac{\sigma}{2\eta} (\cos \thetad - \cos \thetaeq). \]
Obviously, the latter term is negative for $\thetad > \thetaeq$ (and positive for $\thetad < \thetaeq$) and, hence, drives the system towards equilibrium. 
\item An important consequence of the CR-GNBC for \emph{quasi-stationary} states is that it defines a functional dependence between the dynamic contact angle and the contact line speed. In fact, setting $\dthetad = 0$ leads to the relation
\begin{align}
\label{eqn:kinematics/empirical_relation_gnbc_v1}
\cacl = \frac{\visc (-\clspeed)}{\sigma} = \frac{\lambda}{\varepsilon} \, (\cos \thetad - \cos \thetaeq),
\end{align}
or, equivalently, to
\begin{align}\label{eqn:kinematics/empirical_relation_gnbc_v2}
- (\beta \varepsilon) \, \clspeed = \sigma (\cos \thetad - \cos \thetaeq).
\end{align}
By comparing \eqref{eqn:kinematics/empirical_relation_gnbc_v2} with \eqref{eqn:linear_contact_angle_model}, we see that the contact line friction parameter can be indentified with the product of the ``bulk friction'' in the Navier slip condition and the width of the contact line region, i.e.\
\begin{align}
\zeta = \beta \varepsilon.
\end{align}
The latter equation has been proposed before by \cite{Blake2015} in the context of the molecular kinetic theory. Physically, it indicates that, within the present modeling framework, there is only \emph{one} friction mechanism that affects both the slip at the solid boundary and the dynamics of the microscopic contact angle. In this sense, the contact line friction $\zeta$ can be understood as the lumped wall friction of the contact region.
\item From \eqref{eqn:theta_dot_relation} we conclude that the stress component $\inproduct{\ngamma}{(\nabla \mathbf{v}) \, \ndomega}$ vanishes at the contact line for quasi-stationary states, i.e.\ for $\dot\theta=0$. Hence, there appears to be ``perfect slip'' at the contact line in that case. Actually, the concepts of the ``apparent slip length'' $\lambda_a$ (see figure~\ref{fig:geometrical_slip}) and the physical slip parameter defined as $\lambda=\eta/\beta$ must be distinguished for the GNBC model. In fact, the uncompensated Young stress is able to reverse the sign of the velocity gradient at the contact line. In this case, fluid particles at the solid boundary may have a larger tangential velocity than fluid particles slightly above the boundary. This situation corresponds to a \emph{negative} apparent slip length (see figure~\ref{fig:geometrical_slip}). It is, however, caused by the uncompensated Young stress in the velocity boundary condition. The physical slip length parameter $\lambda$ is still positive and finite in all cases.
\begin{figure}
\centering
\includegraphics[width=0.5\columnwidth]{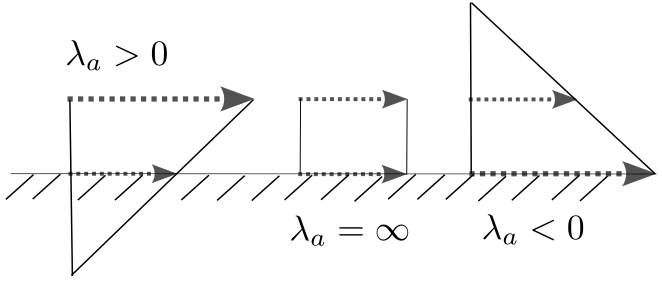} 
\caption{Different cases for the apparent slip length $\lambda_a$: positive, perfect and negative slip (reference frame with $\uwall=0$).}\label{fig:geometrical_slip}
\end{figure}
\item Note that \eqref{eqn:kinematics/theta_evolution_gnbc} can be rephrased as a generalized mobility law of the form 
\begin{align}
\clspeed = f(\thetad, \dthetad).
\end{align}
Therefore, the contact line speed depends on the contact angle $\thetad$ but also on its rate-of-change $\dthetad$ which, in turn, can be computed from $\nabla \mathbf{v}$ (see \cite{Fricke2019}). In this sense, the contact line speed in the GNBC model depends on the flow field in the vicinity of the contact line.
\item Moreover, the GNBC can be understood as an \emph{inhomogeneous} Robin condition for the velocity. Hence, the GNBC enforces a flow whenever $\thetad \neq \thetaeq$. In contrast to the standard \revgen{Navier slip} model, the GNBC model is able to describe the relaxation process of the contact angle.
\end{enumerate}

\subsection{The CR-GNBC thin film equation}\label{subsection:thin_film}
We now derive the thin film equation for the CR-GNBC and compare it with other known thin film equations in the context of dynamic contact lines. The coordinate system is shown in figure~\ref{fig:notation}. Under the thin film assumption, we consider that the pressure remains constant along the $y$-axis and that the Laplace pressure jump across the interface can be expressed as \[\Delta p = - \dfrac{\sigma H^{''}}{(1+H^{'^2})^{3/2}} \approx - \sigma H^{''}. \]
The $x$-momentum equation 
\begin{equation}
    -\frac{1}{\rho} \frac{\partial p}{\partial x} + g + \frac{\visc}{\rho} \frac{\partial^2 v}{\partial y^2} = 0
    \label{eq:x-momentum_thin_film}
\end{equation}
is supplemented by a free surface condition \[\dfrac{\partial v}{\partial y} (x,H) = 0. \] Since we are in the reference frame of the contact line, we can write the CR-GNBC as 
\begin{equation}
    v(x,0) + \lambda \dfrac{\partial v}{\partial y} (x,0) = \uwall [1 - \dirac],
\end{equation}
where $\dirac$ is the smoothed Dirac function that will be defined in \eqref{eq:GNBC_bell}. For further details on the formulation in the reference frame of the contact line, we refer to~\cite{Kulkarni2023}.
Given the contact line boundary condition, the free surface condition and the Laplace pressure jump, the $x$-momentum equation \eqref{eq:x-momentum_thin_film} can now be written as an ordinary differential equation in terms of $H$ 
\begin{equation}
    H^{'''} + \frac{1}{\caplength^2} = - \frac{3 \cawall \left(1- \dirac \right)}{H(H-3\lambda)} - \frac{-3 \visc Q}{\sigma H^2 (H-3\lambda)},
\end{equation}
where $\caplength$ is the capillary length and $Q = -\int_{0}^{H} v \, ds$ is the total flux. Assuming steady state, where $Q=0$ and $H^{'''} \gg 1/\caplength ^2$ in the vicinity of the contact line, we obtain the CR-GNBC thin film equation 
\begin{equation}
    H^{'''} =  \dfrac{3 \cawall \left(\tanh^2 \left(\frac{x}{\varepsilon}\right) \right)}{H(H-3\lambda)}.
    \label{eq:reduced_thin_film_GNBC}
\end{equation}
From \eqref{eq:reduced_thin_film_GNBC}, we can see that for $x \ll \varepsilon$, $H^{''}$ does not diverge and approaches a constant value at the contact line ($x=0$). Hence, the equation is singularity-free. Our CR-GNBC model can therefore be viewed as Navier slip with a \textit{smoothening well} of width $\varepsilon$ around the contact line where the uncompensated Young stress acts.
A comparison of thin film equations from the literature and their respective smoothness is presented in Table~\ref{tab:thin_film}.

\begin{table} 
    \centering
    \caption{Thin film equations for various contact line boundary conditions.}
    \begin{tabular}{ccc}
        \hline
        \textbf{Boundary condition} & \textbf{Thin film equation form} & \textbf{Smoothness} \\
        \hline
        No slip: $v=\uwall$ & $H^{'''} = \dfrac{1}{H^2}$  & Singular $H^{'}$ 
        \\
        & \citet{Duffy1997ATD} & (angle singularity) \\
        \hline
        Navier slip: $v - \lambda \dfrac{\partial v}{\partial y}=\uwall$ & $H^{'''} = \dfrac{3 \cawall}{H^2 + 3 \lambda H}$  & Singular $H^{''} \sim \log x$   \\
        & \citet{Eggers_PRL_2004} & (curvature singularity)\\
        \hline
        Super-slip: $v - \lambda^2\dfrac{\partial^2 v}{\partial y^2}=\uwall$ & $H^{'''} = \dfrac{\cawall}{H^2 + \lambda ^2}$  & Regular $H^{''}$   \\
        & \citet{hocking_2001} & (singularity-free)\\
        \hline
        Super-slip: $v - \lambda \frac{\partial v}{\partial y} - \lambda^2 \frac{\partial^2 v}{\partial y^2}=\uwall$ & 
        $H^{'''} = \dfrac{\cawall}{ {H^2}/{3}  + \lambda H + \lambda^2}$ 
         & Regular $H^{''}$  \\
        & \citet{devauchelle_josserand_zaleski_2007} & (singularity-free)\\
        \hline 
        CR-GNBC &   $ H^{'''} =  \dfrac{3 \cawall (\tanh^2 \frac{x}{\varepsilon})}{H(H-3\lambda)}$  & Regular $H^{''}$  \\
        & current paper & (singularity-free)\\
        \hline 
        \label{tab:thin_film}
    \end{tabular}
\end{table}

%% file: conclusion.tex
To summarize, we have developed an implementation of the Contact Region Generalized Navier Boundary Condition (CR-GNBC) in a geometrical Volume-of-Fluid method. In this method, the dynamic contact angle is not prescribed but is controlled by kinematics through the velocity boundary condition. This is achieved by reconstructing the contact angle at the boundary using the interface normal and the curvature one cell layer away from the boundary. We validate the resulting free contact angle method by studying the interface advection problem in the presence of a moving contact line in Section~\ref{subsection:validation-kinematics}. In the present approach, the uncompensated Young stress is distributed over a characteristic width $\varepsilon$, that is defined independently of the mesh size. Using the kinematic evolution equation of the dynamic contact angle \eqref{eqn:contact-angle-evolution-equation}, we show rigorously that the solution obeys the GNBC law \eqref{eqn:kinematics/theta_evolution_gnbc}, if the solution has a $\mathcal{C}^1$-regularity up to the contact line. Indeed, we show in Section~\ref{sec:Results} that the weak singularity at the contact line is removed in the GNBC model with finite $\varepsilon$. We find a mesh-converging curvature at the contact line (see figure~\ref{fig:curvature_singularity_SLIP_GNBC}) and the numerical solution satisfies the GNBC law in a quasi-stationary state (i.e.\ for $\dthetad=0$). These results are consistent with the recent findings of \cite{Kulkarni2023} who demonstrated that this model indeed shows a local $\mathcal{C}^2$-regularity at the contact line. As expected from kinematics, the tangential stress component goes to zero at the contact line in quasi-stationary states (see figure~\ref{fig:shear_stress_GNBC_SLIP_compare}). In this sense, we observe perfect apparent slip at the moving contact line. A natural follow-up of this work would be to extend the CR-GNBC to non-flat surfaces.


\secondrev{It is also worth noting that VOF and phase-field methods serve complementary roles in contact line modeling. Phase-field methods naturally capture diffuse interface physics at smaller scales, while VOF targets sharp-interface macroscopic simulations where computational efficiency is essential. Systematic benchmarks comparing VOF, phase-field, and molecular dynamics for sheared nanodroplets have been performed by~\citet{Lacis2020,Lācis_Pellegrino_Sundin_Amberg_Zaleski_Hess_Bagheri_2022}, showing that both approaches can match MD results when properly calibrated, albeit through different regularization mechanisms. A direct comparison between the present CR-GNBC formulation and phase-field methods would constitute an interesting direction for future work.}
\\
\\ 
We now discuss in detail the implications and scope of the specific developments achieved in this paper. \\
\begin{enumerate}
\item[{\bf 1.}]  {\it Development of the free contact angle method}\\
We have developed a method that allows us to transport the contact angle in a kinematically consistent manner. This is a major difference to the traditional approaches of imposing constant contact angle or an angle based on a mobility law. A mobility law relates the contact angle with the contact line velocity and other fluid properties like the viscosity ratio, surface tension, surface roughness etc. Vast literature already exists on many of such mobility laws \citep{Xia_Steen_2018,Jacco_review,LUDwicki_Review}. 
In steady-state wetting, where the contact angle remains constant over time, there is a natural inclination to impose a constant contact angle. At this stage, we have made the hypothesis that a constant contact angle exists at nanoscopic scales in steady-state wetting processes. The next question that arises is how to determine the value of this contact angle. Typically, this is decided by solving the Stokes flow equation while assuming a constant contact angle and predicting the interface shape as a function of the capillary number, capillary length, and the contact angle. Several well-known relations exist, such as the Cox law \citep{Cox1986,Voinov1977} and the generalized Cox-Voinov law with the slip boundary condition of \cite{Chan_Kamal_Snoeijer_Sprittles_Eggers_2020}. Mathematically, a simple mobility law is written as $U_{CL} = f (\theta)$, or in an inverse form $\theta = g(U_{CL})$. 
However, one could define generalized mobility laws such that $U_{CL} = f (\theta , \Dot{\theta} , \Ddot{\theta} , ...)$, or in an inverse form $\theta = g(U_{CL} , \dfrac{\partial u}{\partial y} , \dfrac{\partial^2 u}{\partial y^2} , ...)$. Note that the generalized version of the second one involves the gradients of the velocity field at the contact line which would in turn include the outer scales. We interpret the CR-GNBC in section \ref{subsec:Delta_func_approx_GNBC_mobility} as one kind of generalized mobility law. Note that unlike with a simple mobility law, we cannot impose a contact angle directly based on the contact line speed. Here, our free-angle extrapolation scheme proves beneficial. Having successfully demonstrated its capability in the CR-GNBC case, future work could explore its applicability in other forms of the generalized mobility law. \\
\item[{\bf 2.}]{\it A grid independent contact region GNBC}\\
Grid independent results are obtained for a fixed $\varepsilon$ and $\lambda$ with varying grid size $\Delta$, given that $\Delta << \varepsilon$ and $\Delta << \lambda$. We have obtained converging results even with $\varepsilon / \Delta = 5$. Obtaining grid-independent results is crucial for predicting the transition Capillary number as a function of $\varepsilon$ and $\lambda$ so that it could have a potential scope of comparison with experiments. Note that the study of \cite{Afkhami2018} was with no-slip boundary condition giving rise to grid-dependent results in the Volume-of-Fluid framework. Such grid dependency is removed by using a Navier slip and resolving the slip length. We have shown that the curvature diverges logarithmically at the contact line for the Navier slip boundary condition. A divergence in curvature implies a divergence in the pressure field making the model physically ill-posed. Unlike the non-integrable stress singularity that results from the no-slip boundary condition, the Navier slip has an integrable singularity and hence grid-independent steady-state results can be found. A logarithmic divergence of curvature is accompanied by convergence of the contact angle in this case. With the CR-GNBC, given a fixed $\varepsilon$, we were able to confirm the finite curvature as predicted by the thin film equation \ref{subsection:thin_film} and \cite{Kulkarni2023} and get a smooth flow field. A resolution as low as five grid points in the contact region was sufficient to get converged results.\\
\item[{\bf 3.}]{\it Shear stress and the GNBC law}\\
This singularity-free behavior of CR-GNBC over the Navier slip can be seen as a result of incorporating the uncompensated Young stress. Without the uncompensated Young stress, the CR-GNBC reduces back to a classical Navier slip. From \cite{Kulkarni2023}, we know that the Navier slip results in a merely continuous velocity field at the contact line. This implies that the shear stress is mathematically not defined at the contact line. Numerical results in figure~\ref{fig:shear_stress_GNBC_SLIP_compare}b show that we have a non-converging spiked behavior of shear stress at the contact line. From the stream-function solution of \citet{Kulkarni2023}, we can show that the shear stress remains bounded up to the contact line while the differentiability for the shear stress is lost at the contact line. We also observe, from figure~\ref{fig:shear_stress_GNBC_SLIP_compare}, that once uncompensated Young stress is added, the shear stress at the contact line goes to zero in a converging and smooth manner. We numerically recover the GNBC law which relates the steady state contact angle value and the velocity at the contact line in the vanishing shear stress limit. This is perfectly in-line with requirement for a smooth flow from \cite{Fricke2018,Fricke2019}. Notably, this smooth behavior of shear stress going to zero in the CR-GNBC happens only within the $\varepsilon$ width, i.e. within the contact region. 
In the derivation of the GNBC from entropy principles \citep{Fricke2020}, we introduced a smoothed uncompensated Young stress as a Dirac function. This smoothed region can be seen as a physical contact region. However, for mathematical coupling of terms, it remains to be seen how the shear stress would behave if we retained a delta function GNBC formulation. That is, considering an uncompensated Young stress in the singular form of a true delta function, and how it would interact with the singularity of the Navier slip. An investigation of this case is left as a future task.\\
\item[{\bf 4.}]{\it Extension to transient regime}\\
In the paper we have dealt with the steady-state flow characteristics of the GNBC. The work should now be extended to incorporate the setups that have a transient contact line behavior. Examples of such setups include nano-scale shear droplet \citep{Lācis_Pellegrino_Sundin_Amberg_Zaleski_Hess_Bagheri_2022} and a spreading drop. \cite{Forced_vs_sponteneous_wetting} showed that despite using the same fluids and solid material, the overall configuration of the system can result in different values of the apparent contact angle even for equal contact line speeds. The forced wetting case of a plunging plate exhibited a different apparent angle than the spontaneous wetting case of a spreading drop, even when the contact line speed is equal to the plate velocity. Thus a simple mobility law cannot capture this dependence. However, the GNBC contains a term representing the rate of change of the contact angle \ref{eqn:kinematics/theta_evolution_gnbc} which makes spontaneous wetting different from the forced wetting and whether CR-GNBC could explain the experimental behavior remains an interesting question. 
The CR-GNBC should also be extended to setups having sustained oscillatory states. The vibrating drop is a famous example of this setup \cite{Xia_Steen_2018}. There have been several models to describe such oscillatory setups, but most of them rely on empirical relations \citep{kistler1993hydrodynamics}. \cite{Stanley_Sakakeeny_Ling} numerically predicted the first and second modal frequencies of a vibrating droplet in two limiting cases (a) a pinned contact line and (b) a free-slip contact line. In reality, the contact line is expected to behave in between these two limits. Whether a fundamental boundary condition like the CR-GNBC could recover the modal frequencies on large scale as well as the contact line hysteresis at micrometer scale as observed by \cite{Xia_Steen_2018}, remains to be tested. 
\\
\end{enumerate}

%% file: outlook-nonlinear-gnbc.tex
As discussed in detail in Section~\ref{sec:gnbc_derivation}, the CR-GNBC in the form
\begin{align}\label{conclusion/gnbc}
-\beta (\mathbf{v}_\parallel - \vuwall) = (\mat{S}\ndomega)_\parallel + \sigma (\cos \thetad - \cos \thetaeq) \, \ngamma \delta^\varepsilon_\Gamma \quad \text{on} \quad \partial\Omega
\end{align}
is obtained as a \emph{linear} closure relation, to render the dissipation integral
\begin{equation}\label{eqn:conclusion/dissipation-integral}
\begin{aligned}
\mathcal{T} = \int_{\partial\Omega} \left( (\mat{S}\ndomega)_\parallel + \sigma (\cos \thetad - \cos \thetaeq) \, \ngamma \delta^\varepsilon_\Gamma \right) \cdot (\mathbf{v}_\parallel-\vuwall) \, dA
\end{aligned}
\end{equation}
non-positive. Since, according to kinematics, the viscous stress contribution vanishes in a quasi-stationary state (see Section~\ref{sec:mathematical_model}), we obtain the dynamic contact angle relation
\begin{align}\label{eqn:conclusion/linear-response-theory}
- \zeta \clspeed = \sigma (\cos \thetad - \cos \thetaeq)
\end{align}
with the contact line friction coefficient $\zeta=\beta \varepsilon$. Notably, equation~\eqref{eqn:conclusion/linear-response-theory} is also found in the Molecular Kinetic Theory (MKT) in the limit of low capillary number (see, e.g., \cite{Blake2015}). However, for higher capillary numbers, the MKT predicts that\footnote{In this case, the average distance and equilibrium frequency of molecular jumps are denoted by $\Lambda$ and $\kappa^0$, respectively. Moreover, $n$ is the number of adsorption sites per unit area, $k_B$ is the Boltzmann constant and $T$ is the absolute temperature; see \cite{Blake2015} for more details.}
\begin{align}\label{eqn:conclusion/full_mkt}
\clspeed=2 \kappa^0 \Lambda \sinh \left[\sigma\left(\cos \thetaeq-\cos \thetad\right) / (2 n k_B T) \right].
\end{align}
Therefore, it is interesting to formulate a closure relation for \eqref{eqn:conclusion/dissipation-integral} that leads to the relation \eqref{eqn:conclusion/full_mkt} in quasi-stationary states. Notice that \eqref{eqn:conclusion/full_mkt} can be linearized for $\clspeed \rightarrow 0$ using $\sinh(x) = x + \mathcal{O}(x^3)$. Hence, the contact line friction coefficient is identified as $\zeta = (n k_B T)/(\kappa^0 \Lambda)$.\\
\\
For simplicity, let us assume that $\vuwall = 0$ in the following (the generalization to $\vuwall \neq 0$ is obvious). To proceed, it is useful to decompose the integral in \eqref{eqn:conclusion/dissipation-integral} into its components normal and tangential to the contact line, according to
\[ \mathbf{v}_\parallel = (\mathbf{v}_\parallel \cdot \ngamma) \, \ngamma  + (\mathbf{v}_\parallel\cdot \tgamma) \, \tgamma. \]
Here, we denote by $\tgamma$ a unit tangent vector to the contact line. We obtain the representation
\begin{equation}\label{eqn:conclusion/dissipation-integral2}
\begin{aligned}
\mathcal{T} = \mathcal{T}_\bot + \mathcal{T}_\parallel
\end{aligned}
\end{equation}
with
\[ \mathcal{T}_\parallel = \int_{\partial\Omega} \left(\tgamma \cdot (\mat{S}\ndomega)_\parallel \right) (\tgamma \cdot \mathbf{v}_\parallel) \, dA \]
and
\begin{align} 
\mathcal{T}_\bot = \int_{\partial\Omega} \left(\ngamma \cdot (\mat{S}\ndomega)_\parallel + \sigma (\cos \thetad - \cos \thetaeq) \, \delta^\varepsilon_\Gamma \right) (\ngamma \cdot \mathbf{v}_\parallel) \, dA.
\end{align}
We are now looking for closure relations to ensure that $\mathcal{T}_\bot \leq 0$ and $\mathcal{T}_\parallel \leq 0$. A general non-linear closure for $\mathcal{T}_\bot$ reads as
\begin{align}
\boxed{\ngamma \cdot (\mat{S}\ndomega)_\parallel + \sigma (\cos \thetad - \cos \thetaeq) \, \delta^\varepsilon_\Gamma = -f(\ngamma \cdot \mathbf{v}_\parallel),}
\end{align}
where the scalar function $f$ satisfies the inequality
\[ x f(x) \geq 0 \quad \forall x \in \RR. \]
Such a closure is consistent with the second law of thermodynamics because it implies that
\[ \mathcal{T}_\bot = - \int_{\partial \Omega} (\ngamma \cdot \mathbf{v}_\parallel) f(\ngamma \cdot \mathbf{v}_\parallel) \leq 0. \]
The linear version of the GNBC is recovered as $f(x)=\beta x$. \secondrev{Analogously, a closure for the component tangential to the contact line can be obtained as}
\begin{align*}
\secondrev{\tgamma \cdot (\mat{S}\ndomega)_\parallel = -f(\tgamma \cdot \mathbf{v}_\parallel).}
\end{align*}
Motivated by \eqref{eqn:conclusion/full_mkt}, a special choice is
\begin{align}\label{eqn:mkt-closure}
f(x) =  b \, \text{arcsinh}(x/a).  
\end{align}
with positive constants $a=2 \kappa^0 \Lambda$ and $b=2 n k_B T$. Clearly, Equation~\eqref{eqn:mkt-closure} reduces by linearization to the original GNBC \eqref{conclusion/gnbc} with $\beta=b/a$ if $\mathbf{v}_\parallel \cdot \ngamma \rightarrow 0$. Since \eqref{eqn:mkt-closure} corresponds to the MKT \eqref{eqn:conclusion/full_mkt} for quasi-stationary states, it may improve the standard GNBC model for higher values of the capillary number. This shall be studied in detail in the future.

%% file: bibliography.bib
@article{BNA2016291,
title = {Vofi — A library to initialize the volume fraction scalar field},
journal = {Computer Physics Communications},
volume = {200},
pages = {291-299},
year = {2016},
issn = {0010-4655},
doi = {https://doi.org/10.1016/j.cpc.2015.10.026},
url = {https://www.sciencedirect.com/science/article/pii/S0010465515004087},
author = {S. Bnà and S. Manservisi and R. Scardovelli and P. Yecko and S. Zaleski}
}

@article{Forced_vs_sponteneous_wetting,
author = {Karim, M. A. and Davis, S. H. and Kavehpour, H. P.},
title = {Forced versus Spontaneous Spreading of Liquids},
journal = {Langmuir},
volume = {32},
number = {40},
pages = {10153-10158},
year = {2016},
doi = {10.1021/acs.langmuir.6b00747},
    note ={PMID: 27643428},

URL = { 
    
        https://doi.org/10.1021/acs.langmuir.6b00747
    
    

},
eprint = { 
    
        https://doi.org/10.1021/acs.langmuir.6b00747
    
    

}

}

@article{Stanley_Sakakeeny_Ling,
    author = {Sakakeeny, Jordan and Ling, Yue},
    title = "{Numerical study of natural oscillations of supported drops with free and pinned contact lines}",
    journal = {Physics of Fluids},
    volume = {33},
    number = {6},
    pages = {062109},
    year = {2021},
    month = {06},
    issn = {1070-6631},
    doi = {10.1063/5.0049328},
    url = {https://doi.org/10.1063/5.0049328},
    eprint = {https://pubs.aip.org/aip/pof/article-pdf/doi/10.1063/5.0049328/15909411/062109\_1\_online.pdf},
}

@article{kistler1993hydrodynamics,
  title={Hydrodynamics of wetting},
  author={Kistler, Stephan F},
  journal={Wettability},
  volume={6},
  pages={311--430},
  year={1993},
  publisher={Marcel Dekker New York}
}

@article{LUDwicki_Review,
    author = "Ludwicki, JM and Kern, VR and McCraney, J and Bostwick, JB and Daniel, S and Steen, PH",
    title = "Is contact-line mobility a material parameter? NPJ Microgravity. ",
    journal = "NPJ Microgravity",
    year = "2022",
    doi = "10.1038/s41526-022-00190-y",
}

@article{Jacco_review,
   author = "Snoeijer, Jacco H. and Andreotti, Bruno",
   title = "Moving Contact Lines: Scales, Regimes, and Dynamical Transitions", 
   journal= "Annual Review of Fluid Mechanics",
   year = "2013",
   volume = "45",
   number = "Volume 45, 2013",
   pages = "269-292",
   doi = "https://doi.org/10.1146/annurev-fluid-011212-140734",
   url = "https://www.annualreviews.org/content/journals/10.1146/annurev-fluid-011212-140734",
   publisher = "Annual Reviews",
   issn = "1545-4479",
   type = "Journal Article",
  }

@article{Xia_Steen_2018, title={Moving contact-line mobility measured}, volume={841}, DOI={10.1017/jfm.2018.105}, journal={Journal of Fluid Mechanics}, author={Xia, Yi and Steen, Paul H.}, year={2018}, pages={767–783}}

@article{Lācis_Pellegrino_Sundin_Amberg_Zaleski_Hess_Bagheri_2022, title={Nanoscale sheared droplet: volume-of-fluid, phase-field and no-slip molecular dynamics}, volume={940}, DOI={10.1017/jfm.2022.219}, journal={Journal of Fluid Mechanics}, author={Lācis, Uǧis and Pellegrino, Michele and Sundin, Johan and Amberg, Gustav and Zaleski, Stéphane and Hess, Berk and Bagheri, Shervin}, year={2022}, pages={A10}}

@article{Chan_Kamal_Snoeijer_Sprittles_Eggers_2020, title={Cox–Voinov theory with slip}, volume={900}, DOI={10.1017/jfm.2020.499}, journal={Journal of Fluid Mechanics}, author={Chan, Tak Shing and Kamal, Catherine and Snoeijer, Jacco H. and Sprittles, James E. and Eggers, Jens}, year={2020}, pages={A8}}

@article{hocking_2001, title={Meniscus draw-up and draining}, volume={12}, DOI={10.1017/S0956792501004247}, number={3}, journal={European Journal of Applied Mathematics}, publisher={Cambridge University Press}, author={Hocking, L. M.}, year={2001}, pages={195–208}}

@article{Eggers_PRL_2004,
  title = {Hydrodynamic Theory of Forced Dewetting},
  author = {Eggers, Jens},
  journal = {Phys. Rev. Lett.},
  volume = {93},
  issue = {9},
  pages = {094502},
  numpages = {4},
  year = {2004},
  month = {Aug},
  publisher = {American Physical Society},
  doi = {10.1103/PhysRevLett.93.094502},
  url = {https://link.aps.org/doi/10.1103/PhysRevLett.93.094502}
}

@Article{Fumagalli2018,
  author    = {Fumagalli, I. and Parolini, N. and Verani, M.},
  journal   = {Journal of Computational Physics},
  title     = {On a free-surface problem with moving contact line: From variational principles to stable numerical approximations},
  year      = {2018},
  issn      = {0021-9991},
  month     = feb,
  pages     = {253--284},
  volume    = {355},
  comment   = {GNBC derivation sharp interface.},
  doi       = {10.1016/j.jcp.2017.11.004},
  publisher = {Elsevier BV},
}

@article{Duffy1997ATD,
  title={A third-order differential equation arising in thin-film flows and relevant to Tanner's Law},
  author={Brian R. Duffy and Stephen K. Wilson},
  journal={Applied Mathematics Letters},
  year={1997},
  volume={10},
  pages={63-68},
  url={https://api.semanticscholar.org/CorpusID:119352924}
}

@article{devauchelle_josserand_zaleski_2007, title={Forced dewetting on porous media}, volume={574}, DOI={10.1017/S0022112006004125}, journal={Journal of Fluid Mechanics}, publisher={Cambridge University Press}, author={Devauchelle, O. and Josserand, C. and Zaleski, S.}, year={2007}, pages={343–364}}

@Article{Afkhami2009,
  author  = {Afkhami, S. and Bussmann, M.},
  journal = {International Journal for Numerical Methods in Fluids},
  title   = {{Height functions for applying contact angles to 3D VOF simulations}},
  year    = {2009},
  issn    = {02712091},
  month   = {nov},
  number  = {8},
  pages   = {827--847},
  volume  = {61},
  doi     = {10.1002/fld.1974},
}

@Article{Afkhami2008,
  author  = {Afkhami, S. and Bussmann, M.},
  journal = {International Journal for Numerical Methods in Fluids},
  title   = {{Height functions for applying contact angles to 2D VOF simulations}},
  year    = {2008},
  issn    = {02712091},
  month   = {jun},
  number  = {4},
  pages   = {453--472},
  volume  = {57},
  doi     = {10.1002/fld.1651},
}

@Article{Afkhami2018,
  author  = {Afkhami, S. and Buongiorno, J. and Guion, A. and Popinet, S. and Saade, Y. and Scardovelli, R. and Zaleski, S.},
  journal = {Journal of Computational Physics},
  title   = {{Transition in a numerical model of contact line dynamics and forced dewetting}},
  year    = {2018},
  issn    = {00219991},
  month   = {dec},
  pages   = {1061--1093},
  volume  = {374},
  comment = {VOF with subgrid scale modeling at the moving contact line?},
  doi     = {10.1016/j.jcp.2018.06.078},
}

@Article{Blake1969,
  author  = {Blake, T.D and Haynes, J.M.},
  journal = {Journal of Colloid and Interface Science},
  title   = {{Kinetics of liquid-liquid displacement}},
  year    = {1969},
  issn    = {00219797},
  month   = {jul},
  number  = {3},
  pages   = {421--423},
  volume  = {30},
  comment = {Original reference for MKT},
  doi     = {10.1016/0021-9797(69)90411-1},
}

@Article{Blake2006,
  author  = {Blake, T. D.},
  journal = {Journal of Colloid and Interface Science},
  title   = {The physics of moving wetting lines},
  year    = {2006},
  issn    = {0021-9797},
  number  = {1},
  pages   = {1--13},
  volume  = {299},
  doi     = {10.1016/j.jcis.2006.03.051},
}

@Article{Bonn2009,
  author  = {Bonn, D. and Eggers, J. and Indekeu, J. and Meunier, J. and Rolley, E.},
  journal = {Reviews of Modern Physics},
  title   = {Wetting and spreading},
  year    = {2009},
  issn    = {0034-6861},
  number  = {2},
  pages   = {739--805},
  volume  = {81},
  doi     = {10.1103/RevModPhys.81.739},
}

@Article{Cox1986,
  author  = {Cox, R. G.},
  journal = {Journal of Fluid Mechanics},
  title   = {The dynamics of the spreading of liquids on a solid surface. {Part 1}. {Viscous flow}},
  year    = {1986},
  issn    = {0022-1120},
  pages   = {169--194},
  volume  = {168},
  doi     = {10.1017/S0022112086000332},
}

@Article{Fullana2020,
  author  = {Fullana, T. and Zaleski, S. and Popinet, S.},
  journal = {The European Physical Journal Special Topics},
  title   = {{Dynamic wetting failure in curtain coating by the Volume-of-Fluid method}},
  year    = {2020},
  issn    = {1951-6355},
  month   = {sep},
  number  = {10},
  pages   = {1923--1934},
  volume  = {229},
  doi     = {10.1140/epjst/e2020-000004-0},
}

@Book{Gennes2004,
  author    = {de Gennes, P.-G. and Brochard-Wyart, F. and Qu{\'{e}}r{\'{e}}, D.},
  publisher = {Springer New York},
  title     = {{Capillarity and Wetting Phenomena}},
  year      = {2004},
  address   = {New York, NY},
  isbn      = {978-1-4419-1833-8},
  doi       = {10.1007/978-0-387-21656-0},
  pages     = {291},
}

@Article{Gerbeau2009,
  author  = {Gerbeau, J.-F. and Leli{\`{e}}vre, T.},
  journal = {Computer Methods in Applied Mechanics and Engineering},
  title   = {{Generalized Navier boundary condition and geometric conservation law for surface tension}},
  year    = {2009},
  issn    = {00457825},
  number  = {5-8},
  pages   = {644--656},
  volume  = {198},
  doi     = {10.1016/j.cma.2008.09.011},
}

@Article{Huh1971,
  author   = {Huh, C. and Scriven, L. E},
  journal  = {Journal of Colloid and Interface Science},
  title    = {Hydrodynamic model of steady movement of a solid/liquid/fluid contact line},
  year     = {1971},
  issn     = {0021-9797},
  number   = {1},
  pages    = {85--101},
  volume   = {35},
  comment  = {The famous Huh and Scriven.},
  doi      = {10.1016/0021-9797(71)90188-3},
}

@Article{Huh1977,
  author          = {Huh, C. and Mason, S. G.},
  journal         = {Journal of Fluid Mechanics},
  title           = {{The steady movement of a liquid meniscus in a capillary tube}},
  year            = {1977},
  issn            = {0022-1120},
  month           = {jul},
  number          = {03},
  pages           = {401--419},
  volume          = {81},
  doi             = {10.1017/S0022112077002134},
  publisher       = {Cambridge University Press},
}

@Article{Lacis2020,
  author          = {Lācis, U. and Johansson, P. and Fullana, T. and Hess, B. and Amberg, G. and Bagheri, S. and Zaleski, S.},
  journal         = {The European Physical Journal Special Topics},
  title           = {{Steady moving contact line of water over a no-slip substrate}},
  year            = {2020},
  issn            = {1951-6355},
  month           = {sep},
  number          = {10},
  pages           = {1897--1921},
  volume          = {229},
  doi             = {10.1140/epjst/e2020-900280-9},
}

@Article{Lukyanov2017,
  author   = {Lukyanov, A. V. and Pryer, T.},
  journal  = {Langmuir},
  title    = {Hydrodynamics of Moving Contact Lines: Macroscopic versus Microscopic},
  year     = {2017},
  number   = {34},
  pages    = {8582--8590},
  volume   = {33},
  doi      = {10.1021/acs.langmuir.7b02409},
}

@Article{Maric2020,
  author  = {Mari{\'{c}}, T. and Kothe, D. B. and Bothe, D.},
  journal = {Journal of Computational Physics},
  title   = {{Unstructured un-split geometrical Volume-of-Fluid methods – A review}},
  year    = {2020},
  issn    = {00219991},
  month   = {nov},
  pages   = {109695},
  volume  = {420},
  doi     = {10.1016/j.jcp.2020.109695},
}

@Book{Pruess2016,
  author    = {Pr{\"u}ss, J. and Simonett, G.},
  publisher = {Birkh{\"a}user},
  title     = {Moving interfaces and quasilinear parabolic evolution equations},
  year      = {2016},
  address   = {Switzerland},
  isbn      = {978-3319276977},
  series    = {Monographs in Mathematics},
  doi       = {10.1007/978-3-319-27698-4},
  price     = {eBook},
  number = {}
}

@Article{Popinet2009,
  author   = {Popinet, S.},
  journal  = {Journal of Computational Physics},
  title    = {{An accurate adaptive solver for surface-tension-driven interfacial flows}},
  year     = {2009},
  issn     = {00219991},
  number   = {16},
  pages    = {5838--5866},
  volume   = {228},
  doi      = {10.1016/j.jcp.2009.04.042},
}

@Article{Popinet2018,
  author  = {Popinet, S.},
  journal = {Annual Review of Fluid Mechanics},
  title   = {{Numerical Models of Surface Tension}},
  year    = {2018},
  issn    = {0066-4189},
  number  = {1},
  pages   = {49--75},
  volume  = {50},
  doi     = {10.1146/annurev-fluid-122316-045034},
}

@Article{Qian2003,
  author   = {Qian, T. and Wang, X.-P. and Sheng, P.},
  journal  = {Physical Review E},
  title    = {Molecular scale contact line hydrodynamics of immiscible flows},
  year     = {2003},
  issn     = {1539-3755},
  number   = {1 Pt 2},
  pages    = {016306},
  volume   = {68},
  comment  = {Important paper about GNBC.},
  doi      = {10.1103/PhysRevE.68.016306},
}

@Article{Qian2006,
  author  = {Qian, T. and Wang, X.-P. and Sheng, P.},
  journal = {Journal of Fluid Mechanics},
  title   = {{A variational approach to moving contact line hydrodynamics}},
  year    = {2006},
  issn    = {0022-1120},
  month   = {oct},
  pages   = {333},
  volume  = {564},
  comment = {GNBC in the diffuse interface context.},
  doi     = {10.1017/S0022112006001935},
}

@Article{Ren2007,
  author  = {Ren, W. and E, W.},
  journal = {Physics of Fluids},
  title   = {Boundary conditions for the moving contact line problem},
  year    = {2007},
  issn    = {1070-6631},
  number  = {2},
  pages   = {022101},
  volume  = {19},
  doi     = {10.1063/1.2646754},
}

@article{Ren2011,

author = {Ren, W. and E, W.},

doi = {10.4310/CMS.2011.v9.n2.a13},

issn = {15396746},

journal = {Communications in Mathematical Sciences},

number = {2},

pages = {597--606},

title = {{Derivation of continuum models for the moving contact line problem based on thermodynamic principles}},

volume = {9},

year = {2011}

}

@article{Ren2011b,
    author = {Ren, W and E, W},
    title = {Contact line dynamics on heterogeneous surfaces},
    journal = {Physics of Fluids},
    volume = {23},
    number = {7},
    pages = {072103},
    year = {2011},
    month = {07},
    issn = {1070-6631},
    doi = {10.1063/1.3609817},
    url = {https://doi.org/10.1063/1.3609817},
}

@Article{Shikhmurzaev1993,
  author  = {Shikhmurzaev, Y. D.},
  journal = {International Journal of Multiphase Flow},
  title   = {The moving contact line on a smooth solid surface},
  year    = {1993},
  issn    = {03019322},
  number  = {4},
  pages   = {589--610},
  volume  = {19},
  doi     = {10.1016/0301-9322(93)90090-h},
}

@Article{Shikhmurzaev2006,
  author  = {Shikhmurzaev, Y. D.},
  journal = {Physica D: Nonlinear Phenomena},
  title   = {Singularities at the moving contact line. {Mathematical, physical and computational aspects}},
  year    = {2006},
  issn    = {01672789},
  number  = {2},
  pages   = {121--133},
  volume  = {217},
  doi     = {10.1016/j.physd.2006.03.003},
}

@Book{Shikhmurzaev2008,
  author    = {Shikhmurzaev, Y. D.},
  publisher = {{Chapman {\&} Hall/CRC}},
  title     = {Capillary flows with forming interfaces},
  year      = {2008},
  address   = {Boca Raton},
  isbn      = {1-58488-748-6},
  doi       = {10.1201/9781584887492},
}

@book{Slattery.1999,
 author = {Slattery, J. C.},
 year = {1999},
 title = {Advanced transport phenomena},
 address = {Cambridge},
 publisher = {{Cambridge University Press}},
 isbn = {9780511800238},
 series = {Cambridge Series in Chemical Engineering},
 doi = {10.1017/CBO9780511800238},
 number = {}
}

@Article{Snoeijer2013,
  author  = {Snoeijer, J. H. and Andreotti, B.},
  journal = {Annual Review of Fluid Mechanics},
  title   = {Moving Contact Lines: {Scales}, {Regimes}, and {Dynamical Transitions}},
  year    = {2013},
  issn    = {0066-4189},
  number  = {1},
  pages   = {269--292},
  volume  = {45},
  doi     = {10.1146/annurev-fluid-011212-140734},
}

@article{Young.1805,
 author = {Young, T.},
 year = {1805},
 title = {An Essay on the Cohesion of Fluids},
 pages = {65--87},
 volume = {95},
 number = {},
 issn = {0261-0523},
 journal = {Philosophical Transactions of the Royal Society of London},
 doi = {10.1098/rstl.1805.0005}
}

@Article{Scardovelli1999,
  author   = {Scardovelli, R. and Zaleski, S.},
  journal  = {Annual Review of Fluid Mechanics},
  title    = {{Direct Numerical Simulation of Free-Surface and Interfacial Flow}},
  year     = {1999},
  issn     = {0066-4189},
  number   = {1},
  pages    = {567--603},
  volume   = {31},
  doi      = {10.1146/annurev.fluid.31.1.567},
}

@Book{Tryggvason2011,
  author    = {Tryggvason, G. and Scardovelli, R. and Zaleski, S.},
  publisher = {Cambridge University Press},
  title     = {{Direct numerical simulations of gas-liquid multiphase flows}},
  year      = {2011},
  isbn      = {9780511975264},
  doi       = {10.1017/CBO9780511975264},
}

@Article{Chen2019,
  author    = {Xianyang Chen and Jiacai Lu and Gr{\'{e}}tar Tryggvason},
  journal   = {Physics of Fluids},
  title     = {Numerical simulation of self-propelled non-equal sized droplets},
  year      = {2019},
  month     = {may},
  number    = {5},
  pages     = {052107},
  volume    = {31},
  doi       = {10.1063/1.5094757},
  publisher = {{AIP} Publishing},
}

@Article{Fricke2018,
  author  = {Fricke, M. and Köhne, M. and Bothe, D.},
  journal = {PAMM},
  title   = {On the Kinematics of Contact Line Motion},
  year    = {2018},
  number  = {1},
  pages   = {e201800451},
  volume  = {18},
  doi     = {10.1002/pamm.201800451},
}

@Article{Fricke2019,
  author  = {Fricke, M. and K{\"{o}}hne, M. and Bothe, D.},
  journal = {Physica D: Nonlinear Phenomena},
  title   = {{A kinematic evolution equation for the dynamic contact angle and some consequences}},
  year    = {2019},
  issn    = {01672789},
  pages   = {26--43},
  volume  = {394},
  doi     = {10.1016/j.physd.2019.01.008},
}

@Article{Fricke2020,
  author  = {Fricke, M. and Mari\'{c}, T. and Bothe, D.},
  journal = {Journal of Computational Physics},
  volume = {407},
  pages = {109221},
  title   = {{Contact} {Line} {Advection} using the geometrical {Volume of Fluid Method}},
  year    = {2020},
  doi     = {10.1016/j.jcp.2019.109221},
}

@Article{Fricke2020a,
  author  = {Fricke, M. and Bothe, D.},
  journal = {The European Physical Journal Special Topics},
  title   = {Boundary conditions for dynamic wetting - {A} mathematical analysis},
  year    = {2020},
  issn    = {1951-6355},
  month   = {sep},
  number  = {10},
  pages   = {1849-1865},
  volume  = {229},
  doi     = {10.1140/epjst/e2020-900249-7},
}

@Article{Jacqmin2000,
  author    = {David Jacqmin},
  journal   = {Journal of Fluid Mechanics},
  title     = {Contact-line dynamics of a diffuse fluid interface},
  year      = {2000},
  month     = {jan},
  pages     = {57--88},
  volume    = {402},
  doi       = {10.1017/s0022112099006874},
  publisher = {Cambridge University Press ({CUP})},
}

@PhdThesis{Fricke2021,
  author = {M. Fricke},
  school = {TU Darmstadt},
  title  = {Mathematical Modeling and Volume-of-Fluid based simulation of dynamic wetting},
  year   = {2021},
  doi    = {10.12921/tuprints-00014274},
}

@Article{Popinet1999,
  author    = {St{\'{e}}phane Popinet and St{\'{e}}phane Zaleski},
  journal   = {International Journal for Numerical Methods in Fluids},
  title     = {A front-tracking algorithm for accurate representation of surface tension},
  year      = {1999},
  month     = {jul},
  number    = {6},
  pages     = {775--793},
  volume    = {30},
  comment   = {Surface Tension for marker and spline.},
  doi       = {10.1002/(sici)1097-0363(19990730)30:6<775::aid-fld864>3.0.co;2-#},
  publisher = {Wiley},
}

@article{Popinet2015,
  author = 	 {S. Popinet},
  title = 	 {A quadtree-adaptive multigrid solver
  for the Serre--Green--Naghdi equations},
  journal = 	 {J. Comput. Phys.},
  year = 	 {2015},
  volume =       {302},
  pages =        {336--358},
  doi = {https://doi.org/10.1016/j.jcp.2015.09.009}
}

@Article{Yamamoto2013,
  author    = {Y. Yamamoto and T. Ito and T. Wakimoto and K. Katoh},
  journal   = {International Journal of Multiphase Flow},
  title     = {Numerical simulations of spontaneous capillary rises with very low capillary numbers using a front-tracking method combined with generalized Navier boundary condition},
  year      = {2013},
  month     = {may},
  pages     = {22--32},
  volume    = {51},
  comment   = {Referenced by Stephane Zaleski. GNBC boundary condition.},
  doi       = {10.1016/j.ijmultiphaseflow.2012.12.002},
  publisher = {Elsevier {BV}},
}

@Article{Yamamoto2016,
  author    = {Yamamoto, Y. and Higashida, S. and Tanaka, H. and Wakimoto, T. and Ito, T. and Katoh, K.},
  journal   = {Physics of Fluids},
  title     = {Numerical analysis of contact line dynamics passing over a single wettable defect on a wall},
  year      = {2016},
  issn      = {1089-7666},
  month     = aug,
  number    = {8},
  volume    = {28},
  doi       = {10.1063/1.4961490},
  publisher = {AIP Publishing},
}

@Article{Shang2018,
  author    = {Xinglong Shang and Zhengyuan Luo and Elizaveta Ya. Gatapova and Oleg A. Kabov and Bofeng Bai},
  journal   = {Computers {\&} Fluids},
  title     = {{GNBC}-based front-tracking method for the three-dimensional simulation of droplet motion on a solid surface},
  year      = {2018},
  month     = {aug},
  pages     = {181--195},
  volume    = {172},
  doi       = {10.1016/j.compfluid.2018.06.021},
  publisher = {Elsevier {BV}},
}

@Article{Yamamoto2014,
  author    = {Y. Yamamoto and K. Tokieda and T. Wakimoto and T. Ito and K. Katoh},
  journal   = {International Journal of Multiphase Flow},
  title     = {Modeling of the dynamic wetting behavior in a capillary tube considering the macroscopic{\textendash}microscopic contact angle relation and generalized Navier boundary condition},
  year      = {2014},
  month     = {feb},
  pages     = {106--112},
  volume    = {59},
  doi       = {10.1016/j.ijmultiphaseflow.2013.10.018},
  publisher = {Elsevier {BV}},
}

@Article{Qian2006a,
  author  = {T. Qian and X.-P. Wang and P. Sheng},
  journal = {Commun. Comput. Phys.},
  title   = {Molecular Hydrodynamics of the Moving Contact Line in Two-Phase Immiscible Flows},
  year    = {2006},
  number  = {1},
  pages   = {1--52},
  volume  = {1},
}

@Book{Marengo2022,
  editor    = {Marco Marengo and Joel {De Coninck}},
  publisher = {Springer International Publishing},
  title     = {The Surface Wettability Effect on Phase Change},
  year      = {2022},
  doi       = {10.1007/978-3-030-82992-6},
}

@Article{Blake2015,
  author    = {T. D. Blake and J.-C. Fernandez-Toledano and G. Doyen and J. {De Coninck}},
  journal   = {Physics of Fluids},
  title     = {Forced wetting and hydrodynamic assist},
  year      = {2015},
  month     = {nov},
  number    = {11},
  pages     = {112101},
  volume    = {27},
  doi       = {10.1063/1.4934703},
  publisher = {{AIP} Publishing},
}

@Article{Thompson1989,
  author    = {Peter A. Thompson and Mark O. Robbins},
  journal   = {Physical Review Letters},
  title     = {Simulations of contact-line motion: Slip and the dynamic contact angle},
  year      = {1989},
  month     = {aug},
  number    = {7},
  pages     = {766--769},
  volume    = {63},
  doi       = {10.1103/physrevlett.63.766},
  publisher = {American Physical Society ({APS})},
}

@Article{Voinov1977,
  author    = {O. V. Voinov},
  journal   = {Fluid Dynamics},
  title     = {Hydrodynamics of wetting},
  year      = {1977},
  number    = {5},
  pages     = {714--721},
  volume    = {11},
  comment   = {Seminal paper by Voinov.},
  doi       = {10.1007/bf01012963},
  publisher = {Springer Science and Business Media {LLC}},
}

@Article{Kawakami2023,
  author    = {K. Kawakami and Y. Kita and Y. Yamamoto},
  journal   = {Computers {\&} Fluids},
  title     = {Front-tracking simulation of the wetting behavior of an impinging droplet using a relaxed impermeability condition and a generalized Navier boundary condition},
  year      = {2023},
  month     = {jan},
  pages     = {105739},
  volume    = {251},
  doi       = {10.1016/j.compfluid.2022.105739},
  publisher = {Elsevier {BV}},
}

@Article{Kulkarni2023,
  author    = {Yash Kulkarni and Tomas Fullana and Stephane Zaleski},
  journal   = {Proceedings of the Royal Society A: Mathematical, Physical and Engineering Sciences},
  title     = {Stream function solutions for some contact line boundary conditions: Navier slip, super slip and the generalized Navier boundary condition},
  year      = {2023},
  month     = {oct},
  number    = {2278},
  volume    = {479},
  comment   = {Stream function paper Yash.},
  doi       = {10.1098/rspa.2023.0141},
  publisher = {The Royal Society},
}

@article{ZHANG2020-JCP,
title = {A level-set method for moving contact lines with contact angle hysteresis},
journal = {Journal of Computational Physics},
volume = {418},
pages = {109636},
year = {2020},
issn = {0021-9991},
doi = {https://doi.org/10.1016/j.jcp.2020.109636},
url = {https://www.sciencedirect.com/science/article/pii/S0021999120304101},
author = {Jiaqi Zhang and Pengtao Yue}
}
